\documentclass[twocolumn,twocolappendix]{aastex63}
\usepackage{amsmath,amstext,natbib}

\received{May 4, 2020}
\accepted{June 11, 2020}

\shorttitle{}
\shortauthors{K. Su et al.}

\begin{document}

\title{Mid-infrared Studies of HD 113766 and HD 172555: Assessing Variability in the Terrestrial Zone of Young Exoplanetary Systems}

\author[0000-0002-3532-5580]{Kate Y.~L.~Su}
\affiliation{Steward Observatory, University of Arizona, 933 N Cherry Avenue, Tucson, AZ 85721, USA}

\author[0000-0003-2303-6519]{George H. Rieke}
\affiliation{Steward Observatory, University of Arizona, 933 N Cherry Avenue, Tucson, AZ 85721, USA}

\author[0000-0001-9834-7579]{Carl Melis}
\affiliation{Center for Astrophysics and Space Sciences, University of California, San Diego, CA 92093, USA}

\author[0000-0003-4393-9520]{Alan P. Jackson}
\affiliation{Centre for Planetary Sciences, University of Toronto at Scarborough, 1265 Military Trail, Toronto, ON M1C 1A4, Canada}
\affiliation{School of Earth and Space Exploration, Arizona State University, 781 E. Terrace Mall, Tempe, AZ 85287, USA}

\author[0000-0002-5083-3663]{Paul S.~Smith}
\affiliation{Steward Observatory, University of Arizona, 933 N Cherry Avenue, Tucson, AZ 85721, USA}

\author[0000-0003-0006-7937]{Huan Y.~A.~Meng}
\affiliation{Steward Observatory, University of Arizona, 933 N Cherry Avenue, Tucson, AZ 85721, USA}

\author[0000-0001-8612-3236]{Andr\'as G\'asp\'ar}
\affiliation{Steward Observatory, University of Arizona, 933 N Cherry Avenue, Tucson, AZ 85721, USA}

\correspondingauthor{Kate Su}
\email{ksu@as.arizona.edu}

\begin{abstract}

We present multi-epoch infrared photometry and spectroscopy obtained with warm {\it Spitzer}, Subaru and SOFIA to assess variability for the young ($\sim$20 Myr) and dusty debris systems around HD 172555 and HD 113766A. No variations (within 0.5\%) were found for the former at either 3.6 or 4.5 $\mu$m, while significant non-periodic variations (peak-to-peak of $\sim$10--15\% relative to the primary star) were detected for the latter. Relative to the {\it Spitzer} IRS spectra taken in 2004, multi-epoch mid-infrared spectra reveal no change in either the shape of the prominent 10 $\mu$m solid-state features or the overall flux levels (no more than 20\%) for both systems, corroborating that the population of sub-$\mu$m-sized grains that produce the pronounced solid-state features is stable over a decadal timescale. We suggest that these sub-$\mu$m-sized grains were initially generated in an optically thick clump of debris of mm-sized vapor condensates resulting from a recent violent impact between large asteroidal or planetary bodies. Because of the shielding from the stellar photons provided by this clump, intense collisions led to an over-production of fine grains that would otherwise be ejected from the system by radiation pressure. As the clump is sheared by its orbital motion and becomes optically thin, a population of very fine grains could remain in stable orbits until Poynting--Robertson drag slowly spirals them into the star. We further suggest that the 3--5 $\mu$m disk variation around HD 113766A is consistent with a clump/arc of such fine grains on a modestly eccentric orbit in its terrestrial zone.

\end{abstract}


\section{Introduction} \label{sec:intro}

Small bodies in our solar system like asteroids, Kuiper-belt objects and comets are the leftover and fragments of planetesimals that failed to form planets. These bodies are known to be present around stars from the detection of circumstellar dust known as a debris disks, which are created as these planetesimals are destroyed. Debris disks offer the opportunity to characterize planetary systems and their evolution from when they emerge from protoplanetary disks into old age \citep{wyatt08,krivov10,matthews14,hughes18}.  These gas-poor disks are composed of dust grains ranging upward in size from $\lesssim$1 $\mu$m, which are continually replenished by sublimation and collisions of parent bodies as the byproduct of planet formation. Sensitive infrared surveys with space telescopes (e.g., {\it IRAS}, {\it Spitzer}, {\it Herschel}, and {\it WISE}\,) have identified thousands of debris disks around mature stars through measurement of the infrared signal when this dust is heated. These observations trace a pattern of development thought to be similar to that of the solar system \citep*{chen_su_xu20}. Common features, such as the co-existence of warm ($\sim$150 K) and cold ($\sim$50 K) dust, suggest underlying order in debris disk structures and illuminate various processes about the formation and evolution of exoplanetary systems \citep{su14}.

A small percentage of stars just beyond the epoch of natal gas-rich disk dispersal show extremely large infrared excesses \citep{balog09,kennedy_wyatt13,meng17}. These systems have hot ($\gtrsim$300 K) terrestrial zone dust levels sufficient to intercept 1--10\% of the star's light and thermally re-emit it in the mid-infrared. The ages of these extreme systems tend to be between 10 and 200 Myr, so the dust is commonly interpreted as the product of terrestrial planet formation (e.g., \citealt{melis10,melis13,meng14}). In this scenario the high dust levels originate in giant impacts analogous to the event that formed the Earth-Moon system (e.g., \citealt{jackson12}).  However, this interpretation is not unique (e.g., \citealt{melis16}). An alternative possibility is that the high dust levels result from transient dynamical clearing after the gas dispersal of regions inhabited by planetesimals, most likely due to the enhanced excitation from adjacent large bodies such as planets. This mechanism might have operated in the past for the outer asteroid belt due to it close proximity to Jupiter \citep{liou_malhotra97}. A third possibility, under appropriate conditions applicable for a massive belt of planetesimals around young early-type stars, is preservation  of a population of fine grains that traditionally would have been thought to have been blown out of the system on short timescales \citep{thebault19}.

In some systems, it has been discovered that the hot dust emission can be highly variable, supporting the terrestrial planet formation hypothesis. The variability provides a unique opportunity to learn about these events and their roles in terrestrial planet formation \citep{melis12}. Successful models to explain these infrared light curves can extract very diagnostic information about the violent collisions such as the impact time, location, and the dominant sizes of the resulting fragments \citep{su19}. To gain a broader insight to the behavior of these extreme debris disks requires that we understand individual systems, including how rapidly the resulting infrared excess rises and fades and which part of the disk is responsible for any variability.  

Here we report multi-epoch, mid-infrared observations of two young debris disks around HD 113766A and HD 172555 to search for infrared variability. Both stars are well studied infrared excess systems that exhibit prominent solid-state features in the mid-infrared, indicative of abundant sub-$\mu$m-sized grains. HD 113766 is a binary system with a projected separation of 1\farcs4 \citep{holden76,fabricius00}. Only the primary (HD 113766A) has the infrared excess (i.e., a debris disk) \citep{meyer01}. Spectra of the combined system indicate a type of F2V \citep{pecaut12} (obtained with a 2\farcs0 slit), while a spectrum of component B, which is $\sim$0.3 mag fainter at $V$, indicates it is F6V \citep{chen11}. This star belongs to the Lower Centaurus-Crux subgroups of Scorpius-Centaurus OB association and hence is $\sim$17 Myr of age  \citep{pecaut12}. HD 172555 is an A7V star, in the $\sim$20 Myr-old $\beta$ Pic moving group \citep{mamajek_bell14}. It is also a binary, but the low-mass companion is at least 2000 au away \citep{torres06} and has no infrared excess \citep{rebull08,riviere-marichalar14}. Although both systems are at a similar age, HD 113766A is much dustier than HD 172555, with an infrared fractional luminosity ($f_d=L_{IR}/L_{\ast}$) of $\sim$2$\times$10$^{-2}$ for the former and $\sim$7$\times$10$^{-4}$ for the latter \citep{mittal15}.

Observations used in this work are reported in Section \ref{sec:sec2}, including new photometry obtained during the warm {\it Spitzer} mission and mid-infrared spectroscopy obtained with Stratospheric Observatory for Infrared Astronomy (SOFIA) and Subaru. We also review all existing infrared observations (published and unpublished) to assess disk variability. We find significant variations in the hot dust around HD 113766A. However, we find that the flux from HD 172555 is dominated by the output of the stellar photosphere at 3.6 and 4.5 $\mu$m, which shows no variations in our data. In Section \ref{sec:sec3}, we discuss the disk structures inferred from spectral energy distribution (SED) models and spatially resolved images, to analyze the implications of the observed variability. The conclusions are given in Section \ref{sec:sec4}.

\section{Observations \& Results} 
\label{sec:sec2}

\begin{figure*}
    \epsscale{1.1}
    \plotone{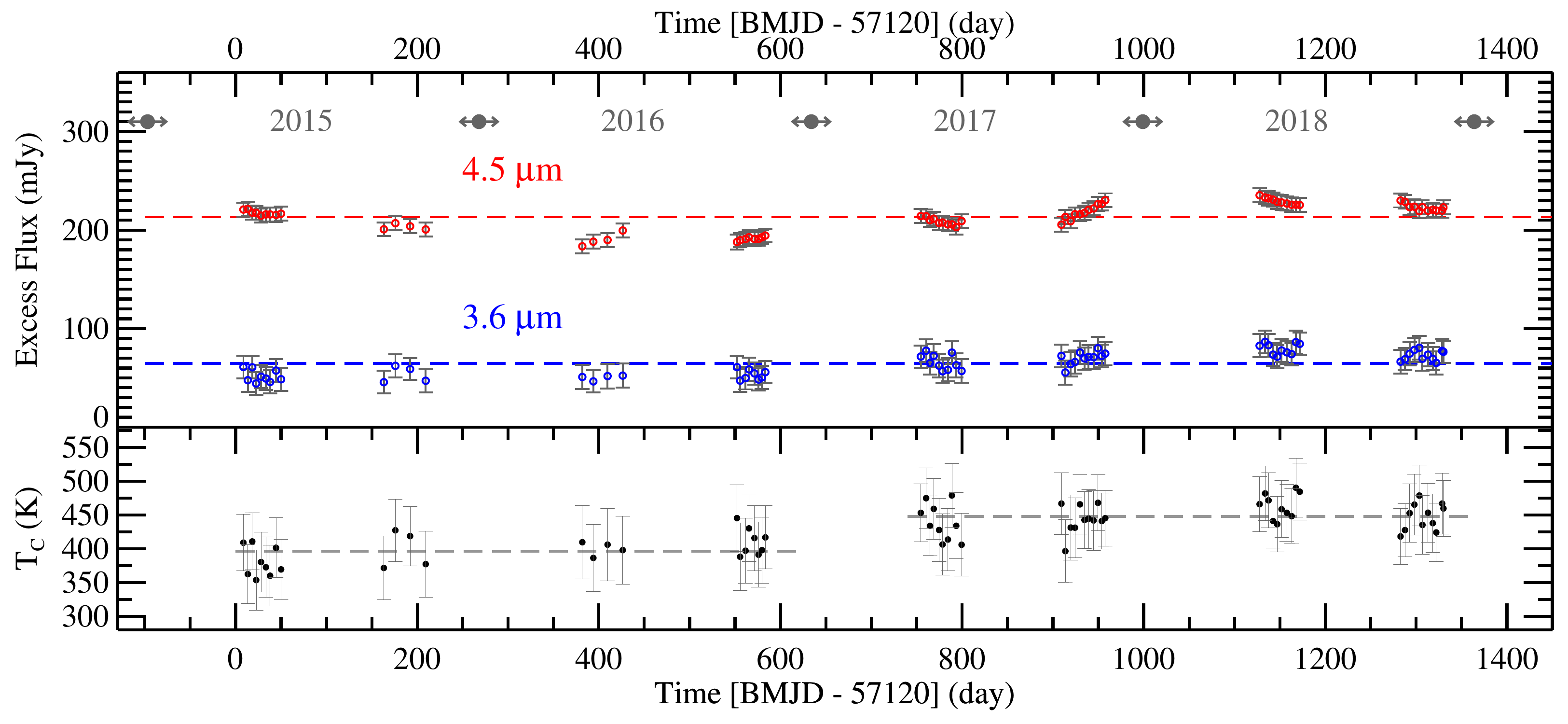} 
    \caption{Time-series {\it Spitzer} observation for the HD 113766 system where the upper panel shows the excess disk emission at 3.6 and 4.5 $\mu$m and the lower panel shows the derived color temperature. In the upper panel, the colored horizontal lines represent the average disk values over the four years of {\it Spitzer} data. In the lower panel, the average color temperatures (396$\pm$24 K and 448$\pm$23 K), using the segments of two-year {\it Spitzer} data are shown as the two horizontal dashed lines. There is a 2$\sigma$ difference in the long-term observed color temperature (details see Section \ref{sec:irac_obs}).  }
    \label{fig:irac_hd113766}
\end{figure*}

\subsection{Warm {\it Spitzer}/IRAC} \label{sec:irac_obs}

Multiple IRAC 3.6 and 4.5 $\mu$m observations were obtained during the {\it Spitzer} warm mission, resulting in a total of 11 Astronomical Observation Requests (AORs) for HD 172555 from PID 90250 (PI: Stapelfeldt) and 11093 (PI: Su), and 67 AORs for HD 113766 from PID 11093. Observations for HD 172555 were performed using sub-array mode to avoid saturation, using a frame time of 0.02 s in a four-point Gaussian dither pattern with medium scale. Only 3.6 $\mu$m data were obtained under PID 90250, but both the 3.6 and 4.5 $\mu$m bands were used under PID 11093. For HD 113766, the full array mode with a frame time of 0.4 s and 10-point cycling dither pattern in medium scale were used. Both objects have two {\it Spitzer} visibility windows per year with each having a length of $\sim$46--52 days. Various observation cadences were used to search for flux variability. A cadence of 10$\pm$2 days was used for HD 172555 under PID 11093 for both visibility windows in 2016, where the flux can be compared to the one obtained in 2013 under PID 90250. For HD 113766, 5$\pm$2 and 15$\pm$2 days were used in individual visibility windows. Details about the observations (AOR Keys, observed date and time) are given in Tables \ref{tab:irac_hd113766} and \ref{tab:irac_hd172555}. 

These data were first processed by the {\it Spitzer} Science Center with the IRAC pipeline S19.2.0. We then performed aperture photometry on the basic calibrated data (BCD) images following the procedure outlined in \citet{meng15} for both full and sub-array data. The resultant photometry is also given in Tables \ref{tab:irac_hd113766} and \ref{tab:irac_hd172555}. The {\it Spitzer} photometry includes contributions from both the star and the dust emission around it.

For HD 172555, we used an aperture 3 pixels in radius and a sky annulus between radii of 12 and 20 pixels (pixel scale of 1\farcs22) with aperture correction factors of 1.112 and 1.113 for 3.6 and 4.5 $\mu$m, respectively. No variability was found within 0.5\% rms in either band between 2013 and 2017. The measurements are also consistent with the expected stellar photosphere (i.e., no infrared excess within a few percent of the stellar photospheric model at these wavelengths). Limited by the optical and near-infrared measurements, the photospheric prediction using stellar models is typically as good as a few percent. We therefore verified the non-infrared excess by comparing signals in the two IRAC bands. Since any residual dust is likely to be at a temperature of $\sim$300--400 K \citep{lisse09}, it would contribute more flux to the 4.5 $\mu$m band than the 3.6 $\mu$m band. However, allowing for the expected color difference between A0V and A7V, the observed fluxes in those bands agree with expectations for the stellar photosphere alone \citep{bohlin11} to within 1.5\%. 

For simplicity, when we refer to the HD 113766 system hereafter, we mean the planetary system around the primary. Because the binary is not completely resolved in the IRAC observations, we used an aperture of 5 pixels (6\farcs1) with a sky annulus of 12--20 pixels (14\farcs6--24\farcs5) to measure the photometry and ensure the flux from both stars is within the aperture. As shown in Table \ref{tab:irac_hd113766}, there is $\sim$5--8\% peak-to-peak variability in the measured total fluxes at both bands. 
 
We determined the observational repeatability by assessing the photometry of field stars in the data. Unfortunately, there is no star within the field of view that is as bright as HD 113766. For fainter field stars (by more than a factor of 10), $\lesssim$2\% repeatability is found at both bands. We conclude that the data on HD 113766 have $\sim$1\% repeatability, similar to other high signal-to-noise measurements obtained in the same program (e.g., \citealt{su19}). This suggests that the 5--8\% variability is significant. The optical photometry (in both $V$ and $g$ bands) from the ASAS-SN network \citep{kochanek17} suggests that the unresolved binary has a combined stellar output stable within 1\%, indicating that the variability comes from the disk around the primary. We determined the combined stellar output to be 728 and 456 mJy in the 3.6 and 4.5 $\mu$m bands, respectively (see Appendix \ref{app:stellar_hd113766} for details). The time-series disk fluxes (at both bands) after subtracting the stellar components and the associated color temperatures from the flux ratio of the two bands\footnote{Following the same procedure as in \citet{su19}, the uncertainties in the color temperatures of the excesses includes 1.5\% of the stellar output. Because the IRAC photometry include both binary components, the representative errors are over estimated.} are shown in Figure \ref{fig:irac_hd113766}. Assuming a standard color between $V$ and the IRAC bands, the two binary components should be of equal brightness for those bands. As a result, correcting for the assumed non-variable emission of component B, the intrinsic variability is about twice that observed, i.e., $\sim$10--15\% (peak-to-peak).  

The four years of the warm {\it Spitzer} data show that the dust emission in the HD 113766 system exhibits non-periodic variability at both 3.6 and 4.5 $\mu$m with peak-to-peak changes at 10--15\% levels (relative to the flux from Component A). There is no strong trend in the overall flux evolution, except that the data taken in the first (last) two years are systematically lower (higher) than the four-year average values  ($F_{IRE,3.6}$=64.7 mJy and $F_{IRE,4.6}$=213.3 mJy) as shown in the upper panel of Figure \ref{fig:irac_hd113766}. This weak trend is also seen in the color temperature: an average of 396$\pm$24 K is derived using the data in the first two years, compared with an average of 448$\pm$23 K in the last two years. There is a 2$\sigma$ positive correlation between the long-term disk flux and color temperature. This suggests that the disk flux variation might be due to changes in dust temperature (i.e., location) if the dust emission is optically thin.

\subsection{SOFIA/FORCAST} \label{sec:sofia_obs}

SOFIA/FORCAST \citep{herter12} observations of HD 172555 were carried out in SOFIA Cycle 4 (Program 04\_0015, PI: Su) on 2016 July 20. Data were obtained in: (1) the grism mode using G111 (covering 8.4--13.7 $\mu$m with a resolution of R=130--260) and G227 (17.6--27.7 $\mu$m, R=110--120), both using the 4\farcs7 slit; and (2) the imaging mode with the F111 ($\lambda_{\rm eff}$=11.1 $\mu$m, $\Delta \lambda$=0.95 $\mu$m) and F242 ($\lambda_{\rm eff}$=24.2 $\mu$m, $\Delta \lambda$=2.9 $\mu$m) filters in the 2-point chopping configuration and a chop throw of 45\arcsec. Observations for HD 113766 were carried out in SOFIA Cycle 5 (Program 05\_0019, PI: Su) on 2017 August 3 using similar instrumental settings. Details about the observations are given in Table \ref{tab:sofia_obs}.

All SOFIA data were processed by the SOFIA Science Center with the pipeline "FORCAST\_REDUX" ver.\ 1\_2\_0 for basic reduction. We used the pipeline-produced science-ready, Level-3 data for further analysis. For imaging data, the Level-3 products are nod-subtracted, merged and flux-calibrated; for spectroscopic data, the Level-3 products also include the optimally extracted, co-added 1-D spectrum (see FORCAST Data Handbook, \url{https://www.sofia.usra.edu/}).

\begin{figure*}
    \epsscale{1.1}
    \plottwo{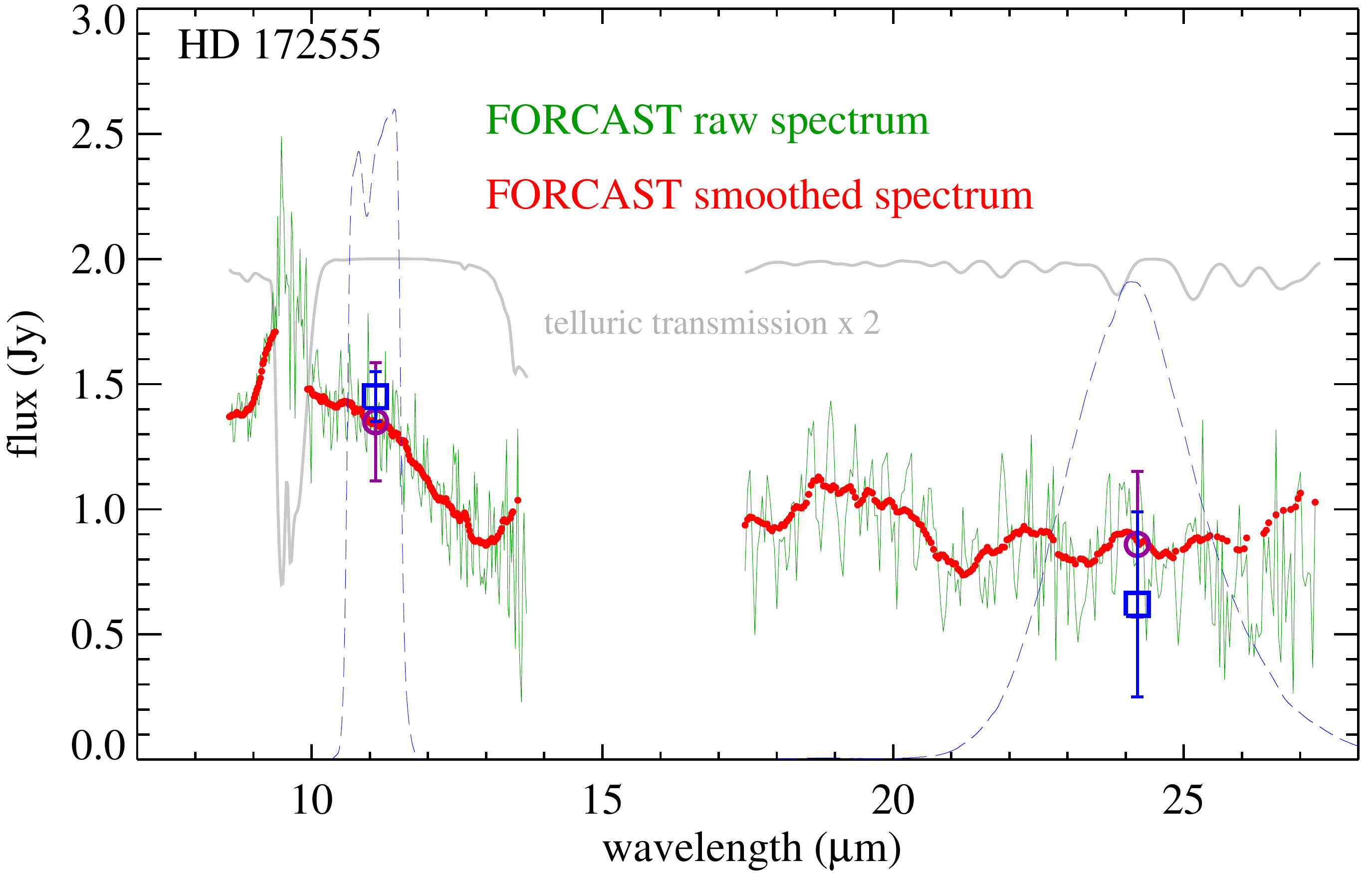}{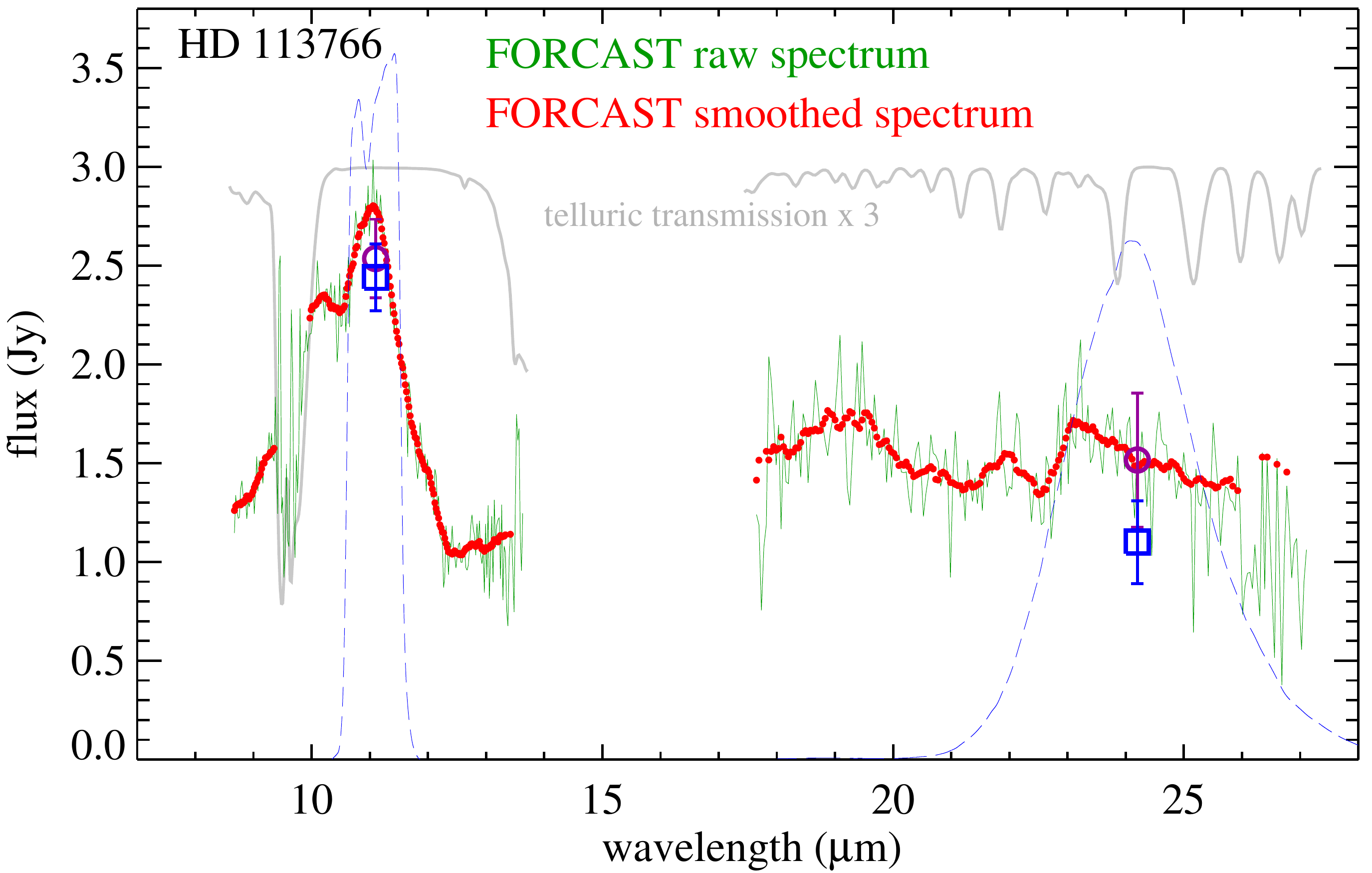} 
    \caption{SOFIA spectra of the HD 172555 (left) and HD 113766 (right) systems. The FORCAST Level-3 raw spectrum is shown as the thin green line, overlaid with smoothed one (red dots, details see Section \ref{sec:sofia_obs}). The thin grey line represents the estimated telluric transmission associated with the data. The data points are discarded when the transmission is below 70\% (particularly the 9.6 $\mu$m region due to atmospheric ozone band). The purple open circles mark the synthesized flux density using the transmission curves (dashed blue lines) of F111 and F242 filters. The photometry obtained in the F111 and F242 images was shown as blue open squares. All error bars are 1$\sigma$ uncertainty. }
    \label{fig:forcast_hd172555_hd113766}
\end{figure*}

Both sources should be point-like in the SOFIA measurements.  Aperture photometry was performed for the imaging data. The centroid of the source was first determined by fitting a 2-D Gaussian function, and then used to center the aperture.  The nominal aperture photometry setting for bright calibrators is an aperture with a radius of 12 pixels ($\sim$4 times the beam defined as the full-width-half-maximum, FWHM), with a sky annulus of 15--25 pixels. However, our sources are much fainter than the bright calibrators (i.e., the photometric noise is dominated by the background), so a smaller aperture setting with an aperture correction is more suitable. For each of the Level-3 image products, we measured the photometry using an aperture setting of 6 pixels and a sky annulus between 6 and 10 pixels. We determined the aperture correction using multiple data sets for four bright flux calibrators observed in the F111 and F242 filters (see Appendix \ref{details_sofia}). The aperture correction factors are 1.127 ($\pm$1.6\% uncertainty) and 1.204 ($\pm$4.8\% uncertainty) for the F111 and F242 filters, respectively.  We estimated the photometric uncertainty as the standard deviation of measurements in 7 non-overlapping apertures in the nominal sky annulus region. For the multiple observations in the F111 filter, we determined an average, weighting by the uncertainty associated with each of the images. An additional 6\% flux calibration uncertainty (FORCAST Data Handbook) was also added in quadrature. The final measured photometry is given in Table \ref{tab:sofia_phot}.

\begin{deluxetable}{ccccc}
\tablenum{1}
\tablewidth{0pc}
\tablecaption{FORCAST Imaging and Spectroscopic Photometry$^a$ \label{tab:sofia_phot}}
\tablehead{
\colhead{HD} & \colhead{F111} & \colhead{F242} & \colhead{G111} & \colhead{G227}
}
\startdata 
172555  &  1.45 $\pm$ 0.13  & 0.62 $\pm$ 0.37 &  1.35 $\pm$ 0.24  & 1.00 $\pm$ 0.29 \\
113766  &  2.44 $\pm$ 0.17  & 1.10 $\pm$ 0.21 &  2.53 $\pm$ 0.20  & 1.51 $\pm$ 0.33
\enddata
\tablenotetext{a}{All fluxes are given in units of Jy.}
\end{deluxetable}

Figure \ref{fig:forcast_hd172555_hd113766} shows the Level-3 spectroscopic results. The wavelength sampling in the default data output is finer than the spectral resolution of the instrument. We first smoothed the raw data (for both grisms) to a uniform resolution of R=80 by including data points that have a signal-to-noise (S/N) greater than 3. Although SOFIA generally flies above 99\% of the Earth's atmosphere, some telluric absorption is inevitable, especially in the 9.6 $\mu$m ozone band. Because telluric calibrators were not observed at the same time  when the target was observed (i.e., at similar flight latitude and airmass), accurate corrections for such absorption are not possible. We excluded the data points for further analysis where the estimated telluric transmission (see the grey curves in Figure \ref{fig:forcast_hd172555_hd113766}) is below 70\%.  Another flux uncertainty in the grism spectra comes when the source is not well centered in the slit during the observation. The main purpose of the imaging data is to evaluate this uncertainty. As shown in Table \ref{tab:sofia_obs}, the imaging data were taken on the same flight as the spectroscopic data, except for HD 113766 where the F242 image was taken three days earlier than the G227 spectrum. We estimated the averaged flux density and its associated uncertainty by summing over the filter transmission curves (also shown in Figure \ref{fig:forcast_hd172555_hd113766}). The resultant synthesized photometry using the spectra is also shown in Table \ref{tab:sofia_phot}. For imaging data, the photometry accuracy ranges from as good as $\sim$7\% in the F111 filter, to much poorer ($\sim$20\% up to 50\%) in the F242 filter. The spectroscopic photometric accuracy is generally in the $\sim$10--30\% range. Considering the uncertainty associated with the data, the flux calibration between the imaging and spectroscopic data agrees within $\pm$1$\sigma$, so no slit loss correction is applied to the FORCAST spectra. We further determined the repeatability of FORCAST grism spectra by assessing multiyear, archival data of bright flux calibrators. A general repeatability of 6\% and 10\% was estimated for the G111 and G227 grism modes (for details, see Appendix \ref{details_sofia}).  In summary, the absolute flux uncertainty associated with the FORCAST grism spectra is in the 6--10\% range.

\subsection{{\it Spitzer}/MIPS Photometry \& SED-mode data} \label{sec:mips_obs}

During the {\it Spitzer} cryogenic mission, multiple observations were obtained with the MIPS instrument for both systems. For HD 172555, there are three data sets available: two of them were in photometry mode with the 24 and 70 $\mu$m channels, and one in the MIPS-SED mode. The first set of photometry was taken on 2004 Apr 06 (AOR Key 3723776, PID 10) and the data were published in \citet{chen06}; the second set was taken on 2008 May 18 (AOR Key 25964288, PID 50316). The MIPS-SED mode data were taken in 2007 Oct 27 (AOR Key 21942528, PID 40679). For HD 113766, two sets of MIPS data were obtained. The MIPS photometry (AOR Key 4789760, PID 84) was taken on 2004 Feb 23, and published in \citet{chen06}. The other set was taken in the MIPS-SED mode (AOR Key 13625856, PID 241) on 2005 Aug 2.

For consistency, we re-reduced all the photometry using an in-house MIPS data pipeline. The images of the sources are point-like in both bands; therefore, we extracted the final values using PSF fitting (for details, see \citealt{sierchio14}). For HD 172555, the MIPS photometry in 2004 is: $F_{24}=865.9 \pm 0.15$ mJy and $F_{70}=226.4 \pm 5.9$ mJy for 24 and 70 $\mu$m, respectively. The MIPS photometry in 2008 is: $F_{24}=865.9 \pm 0.15$ mJy and $F_{70}=226.9 \pm 3.3$ mJy for 24 and 70 $\mu$m, respectively. {\it Herschel}/PACS photometry was also obtained at 70 $\mu$m in 2010 (see Section \ref{sec:ancillary} and Table \ref{tab:sed}), with a result of $217.30 \pm 7.27$ mJy.  That is, there is no flux variation among 2004, 2008, and 2010 within the instrumental calibration levels (1\% and 5\% at 24 and 70 $\mu$m, respectively). For HD 113766 (for both A and B components), the MIPS photometry from 2004 is: $F_{24}=1459.0 \pm 14.6$ mJy and $F_{70}=388.2 \pm 22.4$ mJy, which are consistent with the values reported by \citet{chen11}. The {\it Herschel}/PACS 70 $\mu$m measurement (Table \ref{tab:sed}) from 2011 is $414.78 \pm 7.62$ mJy, again consistent within the expected errors. 

The MIPS-SED mode provides a low-resolution ($R$=15--25) spectrum from 55 to 95 $\mu$m, which overlaps with the MIPS 70 $\mu$m channel and provides additional spectral slope information on the disk SEDs. We extracted the raw MIPS-SED mode data from the {\it Spitzer} archive and reduced them with the MIPS instrument team in-house pipeline (details see \citealt{su15}). Synthesized photometry for the MIPS 70 $\mu$m channel was also computed using these MIPS-SED mode spectra. For HD 172555, the 2007 synthesized photometry is 252$\pm$25 mJy, consistent with the MIPS 70 $\mu$m photometry obtained in both 2004 and 2008. For HD 113766, the 2005 MIPS-SED synthesized photometry is 371$\pm$37 mJy, consistent with the MIPS 70 $\mu$m photometry taken in 2004. In summary, there is no flux variation (within 1\% and 5\% of the measurements) using the MIPS instruments. These data will be discussed together in Section \ref{sec:diskstructure}.

\begin{figure}
    \epsscale{1.1}
    \plotone{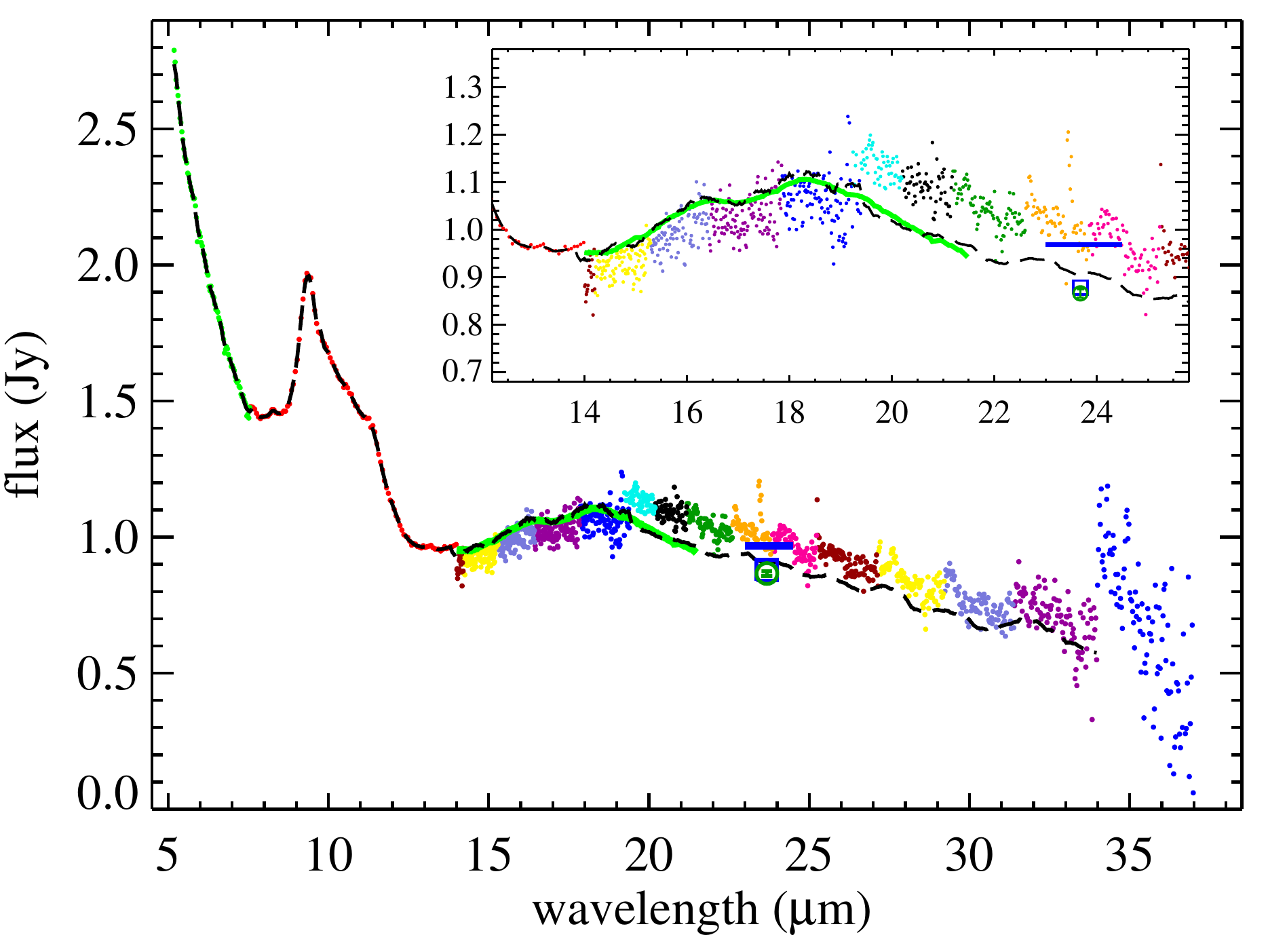} 
    \caption{{\it Spitzer} IRS spectrum of the HD 172555 system. The published 2004 IRS spectrum is shown as small dots with different colors representing different orders. The 2007 LL2 spectrum is shown as thick green line (covering 14--21.5 $\mu$m). The 2004 high-resolution model data were scaled to match the 2007 LL2 data and joined smoothly with the 2004 low-resolution data. A smoothed version of the 2004 spectrum is shown as the dashed black line. For comparison, the synthesized MIPS 24 photometry is 0.969 Jy (thick blue bar) before scaling, but 0.878 Jy (blue open square) after scaling. The average MIPS 24 $\mu$m photometry, 0.866$\pm$0.009 Jy is shown as a dark green open circle for comparison. }
    \label{fig:irs_hd172555}
\end{figure}

\subsection{{\it Spitzer}/IRS} \label{sec:irs_obs}

Both systems have {\it Spitzer} IRS observations. For the HD 113766 system, the measurements were published by \citet{chen06} and \citet{lisse08} along with detailed mineralogical models. We simply used the published spectrum for later analysis. For the HD 172555 system, there were two sets of IRS observations. The first one was taken on 2004 Mar 22 using the IRS Mapping mode (AOR Key 3563264, PID 2) with a combination of low- and high-resolution modules. This data set has been published by \citet{chen06} and \citet{lisse09}. The second set was taken in 2007 Nov 06 using IRS Staring mode (AOR Key 24368384, PID 1446) only in the second order low-resolution module (LL2). We extracted this LL2 spectrum from the CASSIS project \citep{cassis_ref} using the optimal extraction. The two spectra (published and unpublished LL2) are shown in Figure \ref{fig:irs_hd172555}. It appears that there is a flux jump between the low- and high-resolution part of the 2004 spectrum, while the 2007 LL2 spectrum joins more smoothly with the 2004 SL data. Given that there is no flux variation in the MIPS photometry within 1\% at 24 $\mu$m (Section \ref{sec:mips_obs}), the flux discrepancy between the IRS spectra taken in 2004 and 2007 is most likely due to calibration issues between different modes. We scaled the 2004 high-resolution mode data to match the 2007 LL2 data (SH module multiplying by 1.04 and LH module multiplying by 0.91), joining them smoothly with the 2004 SL data. We further smoothed the re-scaled 2004 high-resolution spectrum to R=60 to match the low-resolution mode; the final 2004 re-scaled, smoothed spectrum is also shown in Figure \ref{fig:irs_hd172555}. 

Because the IRS spectrum covers the entire band of the MIPS 24 $\mu$m channel, we further compute the synthesized MIPS 24 photometry using the published and re-scaled 2004 spectra. The derived MIPS 24 $\mu$m photometry is 0.969 and 0.878 Jy, before and after re-scaling, respectively. The re-scaled synthesized photometry is within 1.4\% of the MIPS 24 $\mu$m photometry, consistent within a few percent as expected for the absolute flux calibration established across all {\it Spitzer} instruments \citep{rieke08,bohlin11}. A similar comparison is done for the HD 113766 system using the IRS spectrum taken in 2004 March. The IRS synthesized MIPS 24 $\mu$m photometry is 1464 mJy, matched well with the MIPS 24 $\mu$m photometry ($F_{24}=1459.0\pm14.6$ mJy).

\subsection{Subaru/COMICS} \label{sec:comics_obs}

COMICS is the Cooled Mid-Infrared Camera and Spectrometer mounted on the Subaru 8.2\,m telescope at Mauna Kea \citep{kataza00_comics,okamoto03_comics}. We obtained COMICS observations of HD 113766 on 2017 January 15. To complement these data and further search for changes in the solid-state features, we also retrieved archival COMICS data of HD 113766, taken on 2006 January 13, from SMOKA \citep{baba02}. Observations were taken in low-resolution, N-band spectroscopic mode with a slit width of 0\farcs4 (R$\sim$170) and a chop throw of 10\arcsec\ in 2017, and with a slit width of 0\farcs33 (R$\sim$250) and a chop throw of 12\arcsec\ in 2006. All data were taken in various short integration times on multiple (2--30) chop pairs depending on the desired S/N. 

Data reduction follows standard high thermal background techniques and is executed through in-house IDL routines. First, each chop pair was differenced and examined for quality. We discarded the chop pairs showing strong residual structures due to either rapidly varying background conditions or the source not reasonably centered on the slit. The chop pairs were then rectified along the spatial axis such that the night sky emission lines run vertically in the array coordinate. A second pass of background removal was done by subtracting the median value of background-only pixels along the spatial axis at each dispersion pixel location. Fully processed chop pairs were then median-combined and the positive and negative spectral beams were extracted via a straight aperture sum. Uncertainties on spectral samples were calculated by determining the rms of background-only pixels and summing that in quadrature for the number of pixels in the aperture and the Poisson noise on the total summed flux in the aperture.

\begin{figure}
    \epsscale{1.1}
    \plotone{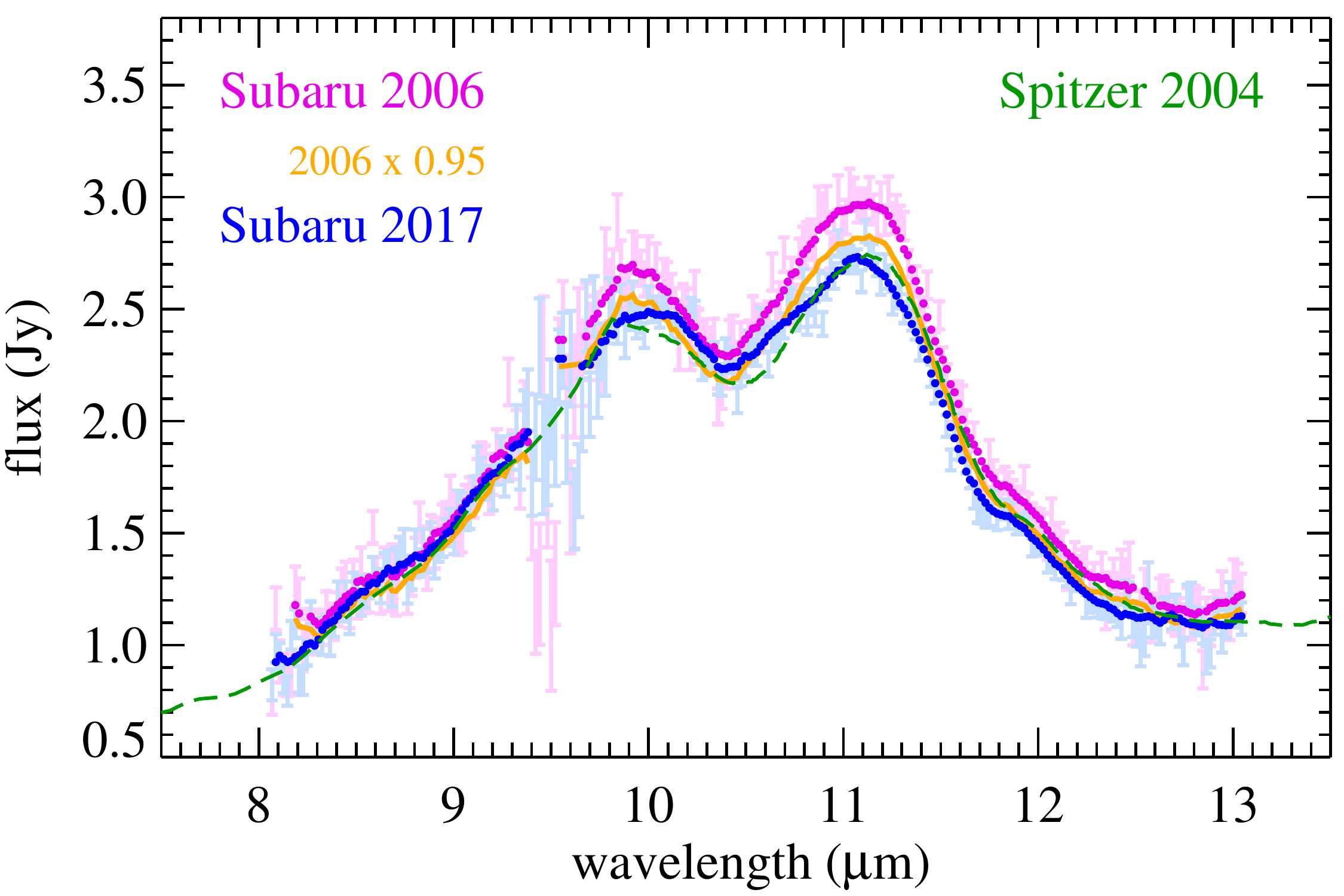} 
    \caption{Subaru/COMICS spectra of HD 113766 taken in 2006 (pink color) and 2017 (blue color). The error bars show 1$\sigma$ uncertainties, excluding the 10\% flux calibration error. Dots represent the smoothed spectra in comparison to the 2004 {\it Spitzer} spectrum (dashed green line). The two COMICS spectra are very similar in shape, and in overall flux (orange line is the 2006 spectrum shifted down by 5\%).  }
    \label{fig:comics_hd113766}
\end{figure}

HD 102964 (K3III) and HR 5029 (K1III) were used as calibrators in 2006 and 2017, respectively. Assuming the calibrators are Rayleigh-Jean-like in 8--13 $\mu$m, telluric correction was performed by dividing each science spectrum by a calibrator spectrum shifted in wavelength to provide the best cancellation of the strong $\sim$9.6\,$\mu$m ozone feature. The wavelength calibration was determined using a low-order polynomial fit to the position of known bright sky emission lines. Individual calibrated science target spectra were then scaled and combined via weighted mean. Flux calibration was performed using the photometry measured from the images of the target and calibrators. We adopted the {\it WISE} W3 measurements for the calibrators, and scaled them to the wavelength of the N-band at which the images were taken. For the 2006 data, both target and calibrator had imaging-only observations. For the 2017 data, no separate imaging-only observation was taken, so we had to rely on the target spectral acquisition images where the star was well-displaced from the slit. The overall flux calibration is accurate at the $\sim$10\% level.

Final spectra showing the 10-$\mu$m range covered by COMICS are shown in Figure \ref{fig:comics_hd113766}. For easy comparison, the calibrated spectra were both smoothed to R=100 after rejecting points with a S/N less than 12. The shape of the 10 $\mu$m feature is very similar between the two spectra taken 11 years apart. The overall flux also agrees well after shifting the 2006 spectrum down by 5\% (within the 10\% flux uncertainty). Furthermore, no significant difference is found between both the COMICS spectra and 2004 {\it Spitzer} IRS spectrum (shown as the dashed green line in Figure \ref{fig:comics_hd113766}).

\begin{deluxetable*}{cllllll}
\tablewidth{0pc}
\tablecaption{SED Measurements\label{tab:sed}}
\tablehead{
\colhead{$\lambda_{\rm eff}$} & \colhead{F$_{\rm tot}$}& \colhead{F$_{\rm IRE}$}  & \colhead{Note}  & \colhead{F$_{\rm tot}$}& \colhead{F$_{\rm IRE}$}  & \colhead{Note} \\
\colhead{($\mu$m)} & \colhead{(mJy)}  & \colhead{(mJy)} & \colhead{}  & \colhead{(mJy)}  & \colhead{(mJy)}  & \colhead{}
}
\startdata 
  & \multicolumn{3}{c}{\underline{HD 172555}}& \multicolumn{3}{c}{\underline{HD 113766}} \\ 
   8.23 &  1450.97$\pm$17.20 &   327.14$\pm$ 22.48 & {\it AKARI}, 2006         &  1308.00$\pm$ 65.00& 1164.42$\pm$ 65.06 & {\it AKARI}, 2006        \\
  11.56 &  1124.87$\pm$15.43 &   545.15$\pm$19.30 & {\it WISE}, 2010          &  1260.96$\pm$11.56 &  1187.18$\pm$11.65 & {\it WISE}, 2010         \\
  12.00 &  1520.00$\pm$60.80 &   981.18$\pm$61.75 & {\it IRAS}, 1983          &  1580.00$\pm$94.80 &  1511.45$\pm$94.81 & {\it IRAS}, 1983         \\
  17.61 &   921.00$\pm$ 19.60 &   667.14$\pm$ 5.08 & {\it AKARI}, 2006         &  1428.00$\pm$50.00 &  1395.82$\pm$50.00 & {\it AKARI}, 2006        \\
  22.09 &   955.79$\pm$19.17 &   793.56$\pm$19.44 & {\it WISE}, 2010          &  1665.57$\pm$18.31 &  1645.05$\pm$18.31 & {\it WISE}, 2010         \\
  23.67 &   865.90$\pm$ 8.66$^{\dagger}$ &   724.62$\pm$ 9.11 & {\it Spitzer}, 2004--2008 &  1459.00$\pm$14.59 &  1441.14$\pm$14.60$^{\dagger}$ & {\it Spitzer}, 2004      \\
  25.00 &  1090.00$\pm$54.50 &   963.25$\pm$54.56 & {\it IRAS}, 1983          &  1800.00$\pm$90.00 &  1783.99$\pm$90.00 & {\it IRAS}, 1983         \\
  60.00 &   306.00$\pm$45.90 &   284.01$\pm$45.90 & {\it IRAS}, 1983          &   622.00$\pm$62.20 &   619.24$\pm$62.20 & {\it IRAS}, 1983         \\
  70.00 &   217.30$\pm$ 7.27 &   201.19$\pm$ 7.28 & {\it Herschel}, 2010      &   414.80$\pm$7.62  &   412.78$\pm$7.62  & {\it Herschel}, 2011     \\
  71.42 &   226.78$\pm$11.34$^{\dagger}$ &   211.31$\pm$11.35 & {\it Spitzer}, 2004--2008 &   388.20$\pm$22.40 &   386.26$\pm$22.40$^{\dagger}$ & {\it Spitzer}, 2004--2008\\
 100.00 &   103.00$\pm$7.93  &    95.16$\pm$ 7.93 & {\it Herschel}, 2011      &   222.50$\pm$7.55  &   221.52$\pm$ 7.55 & {\it Herschel}, 2012     \\
 160.00 &    59.21$\pm$19.69 &   56.19$\pm$ 19.69 & {\it Herschel}, 2010--2011      &    96.13$\pm$ 10.78 &    95.75$\pm$ 10.78 & {\it Herschel}, 2011--2012    \\
1300.00 &     0.11$\pm$ 0.03 &    0.07$\pm$ 0.03  & ALMA, 2012          &     0.62$\pm$0.08  &     0.61$\pm$ 0.08 & ALMA, 2014         
\enddata
\tablecomments{Photometry is given as the total (star+disk) flux (F$_{\rm tot}$) and the excess flux  (F$_{\rm IRE}$) after photospheric subtraction. The note column gives the source of photometry and the rough date that it was taken. $^{\dagger}$ The MIPS photometry uncertainty is limited by the calibration (1\% and 5\% at 24 and 70 $\mu$m, respectively), the reported uncertainty includes the calibration error added in quadrature. }
\end{deluxetable*}

\begin{figure*}
    \epsscale{1.1}
    \plottwo{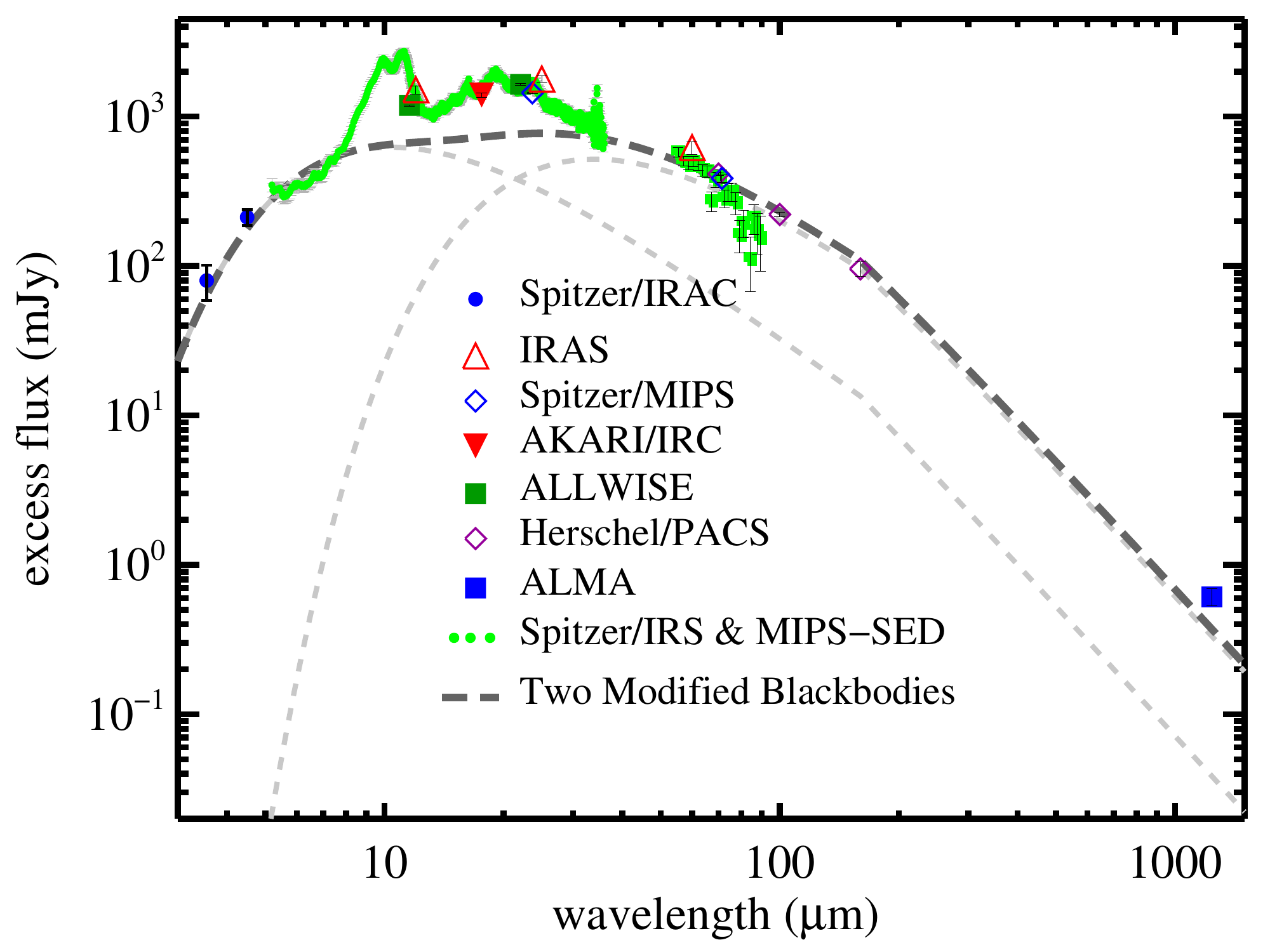}{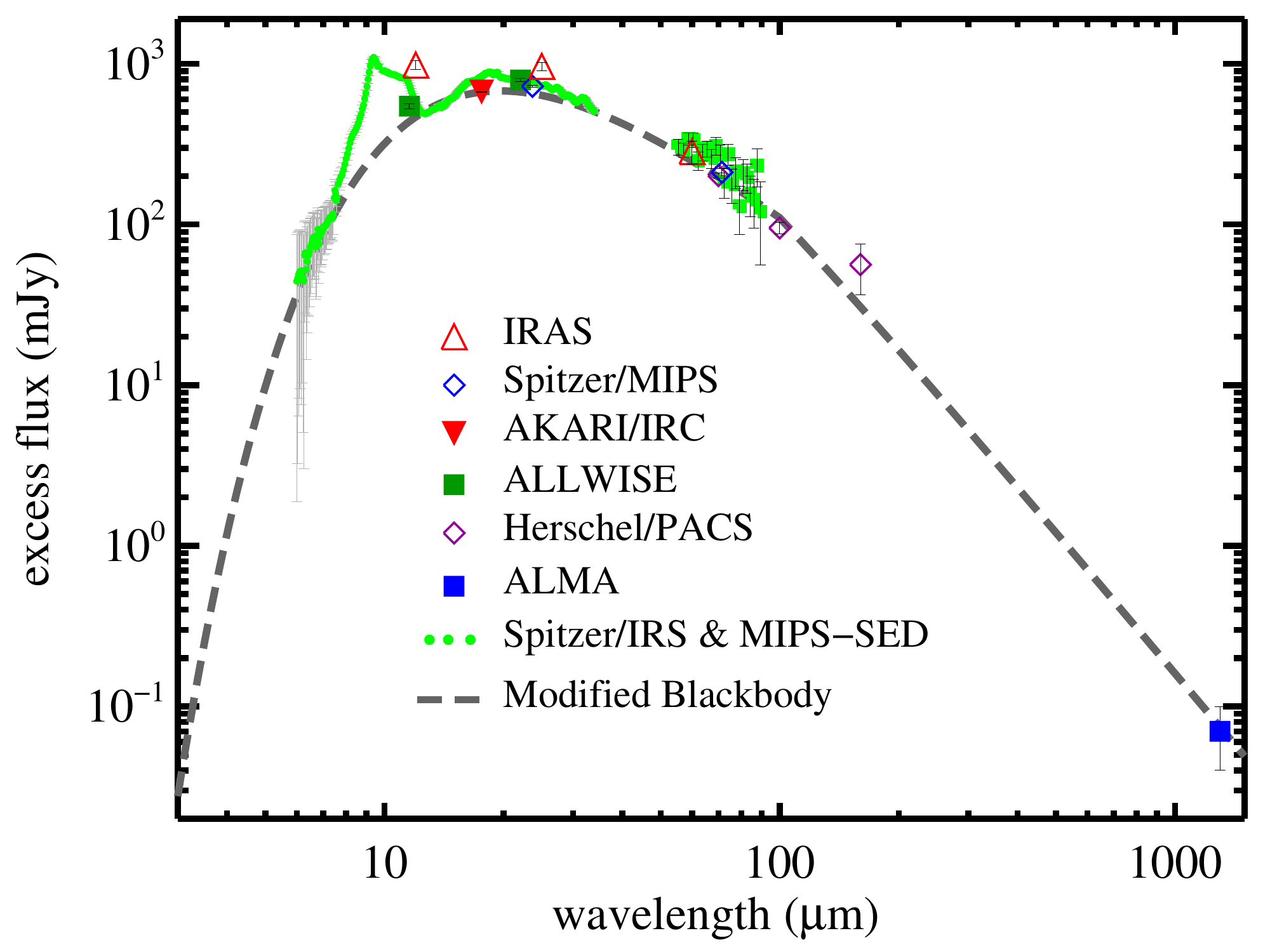} 
    \caption{Spectral energy distribution of the debris emission in the HD 113766 (left) and HD 172555 (right) systems. Observed data are shown as various color symbols given on the plot. The associated error bars are the 1$\sigma$ uncertainty, except for the IRAC measurements for HD 113766A where the error bars reflect the range of variations. For the HD 113766 system, the disk continuum (dark grey, long-dashed line) can be described as the the sum of two modified blackbody functions (light grey, dashed lines) for the inner and outer parts, while only a single such function is needed for the HD 172555 system. }
    \label{fig:sed_hd172555_hd113766}
\end{figure*}

\subsection{Other Ancillary Data} \label{sec:ancillary}

In addition to the data described in the previous subsections, we also collect all available infrared and millimeter photometry for both systems. Although both systems are saturated in the two shortest {\it WISE} bands (W1 and W2), the data from the two longer bands (W3 and W4) are not affected. We adopt the  measurements given in the ALLWISE catalog \citep{cutri14}. We also include {\it AKARI}/IRC and {\it IRAS} measurements\footnote{These sources are not detected by {\it AKARI}/FIS nor IRAS 100 $\mu$m.}. For IRAS, we adopt the measurements from the Faint Source Catalog (FSC). {\it Herschel}/PACS observations have been reported in the literature for both systems \citep{riviere-marichalar14,olofsson13}, but those measurements were based on an early reduction pipeline and calibration. Since both sources are not resolved by {\it Herschel}, we adopt the values from the {\it Herschel} Point Source catalog \citep{marton17} where all available data were combined and processed with the final data calibration and pipeline. Table \ref{tab:sed} lists all the valid observations along with the years they were obtained.  For the measurement uncertainty, we used both the quoted statistical error and the rms of the blank sky region added in quadrature. Finally, both systems have ALMA Band 6 (1.3 mm) observations (\citealt{lieman-sifry16} for HD 113766, and Matr\`a et al.\ in prep. for HD 172555). We further discuss the millimeter observations in Section \ref{sec:diskstructure}.  The collected  measurements are listed in Table \ref{tab:sed} along with the excess emission at those wavelengths after subtracting the stellar component. The disk SEDs (i.e., excess emission) are shown in Figure \ref{fig:sed_hd172555_hd113766}. 


\begin{figure*}
    \epsscale{1.1}
    \plottwo{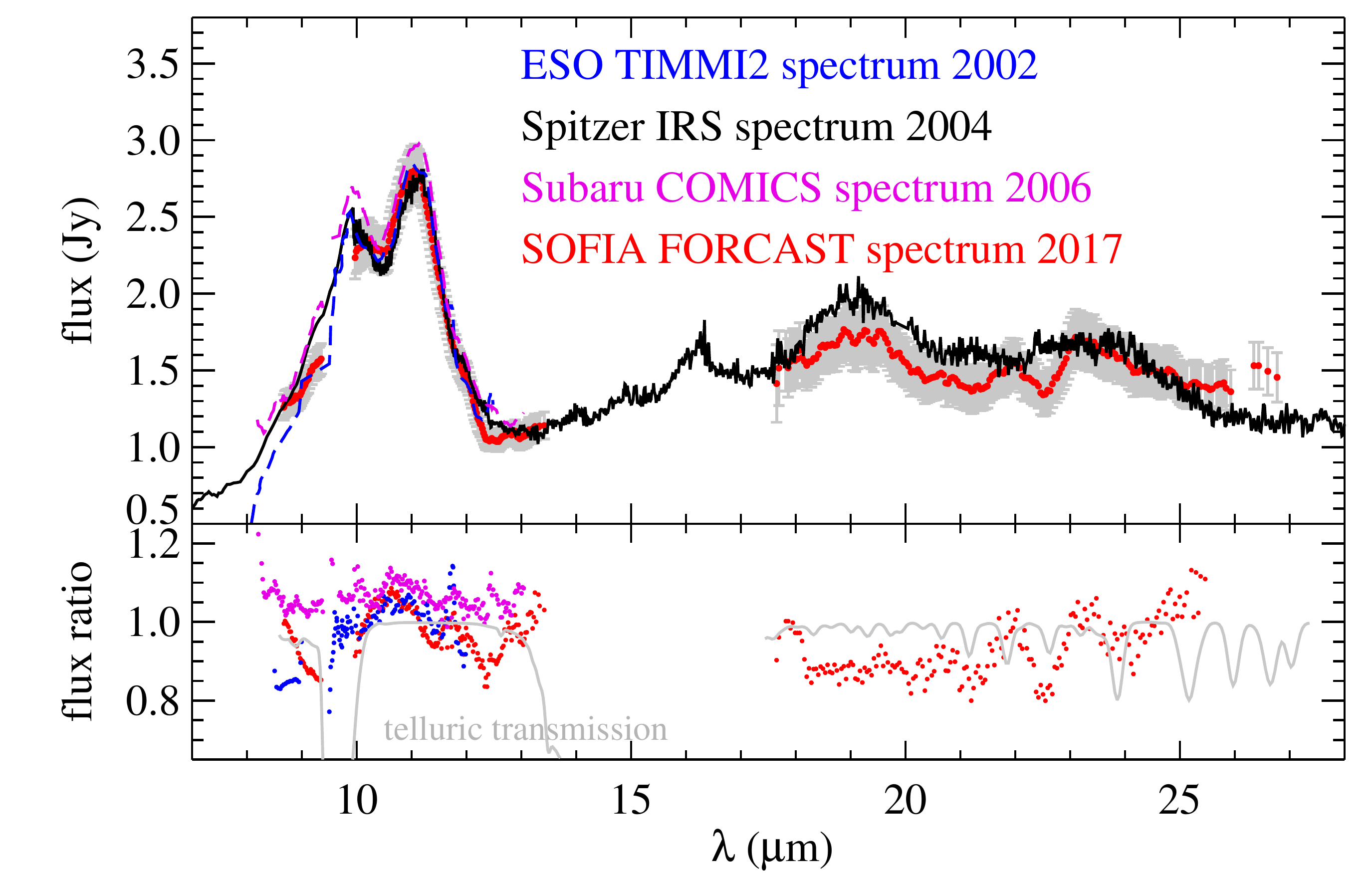}{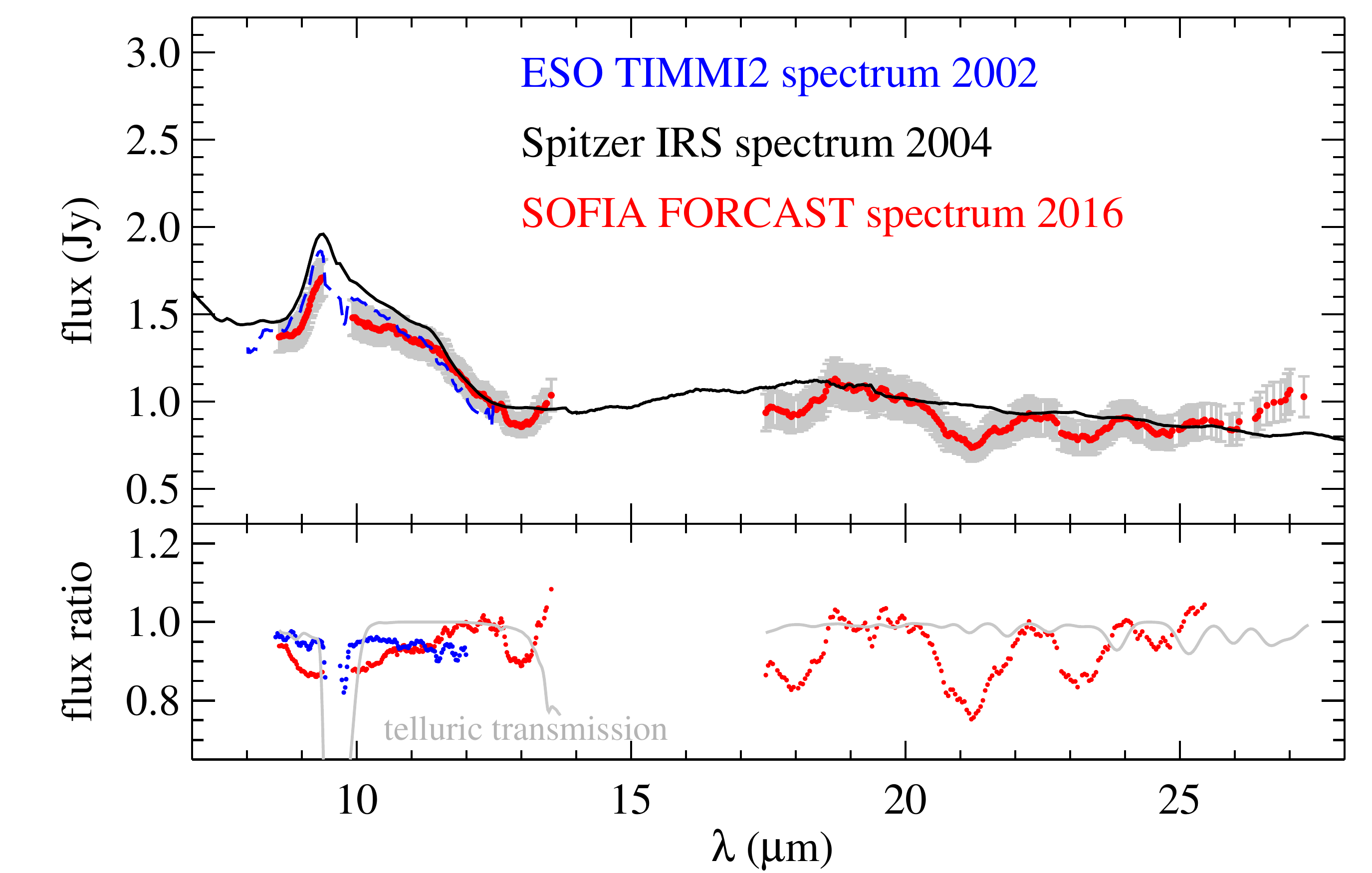} 
    \caption{Comparison of mid-infrared spectra obtained over $\sim$15 year span: left for the HD113766 system and right for the HD 172555 system. The upper panel of the plots shows the spectra in flux level, while the lower panel shows the flux ratio relative to the {\it Spitzer} spectrum taken in 2004. The grey bars show the 1 $\sigma$ uncertainty for the SOFIA spectrum (6\% for G111 and 10\% for G227). In the 10 $\mu$m region covering the prominent solid-state features from sub-$\mu$m grains, no significant (more than 20\%) change is found. Variation at $\lesssim$10\% level might be present, but cannot be confirmed due to uncertainties across different telescopes/instruments. 
    \label{fig:comp_midir_spectra} }
\end{figure*}

\subsection{Variability in Mid-infrared Spectroscopy}
\label{sec:variability}

Because {\it Spitzer} IRS spectra cover several mid-infrared bands in various infrared space missions, one can, in principle, use synthesized photometry to search for variability (similar to the comparison between MIPS 24 $\mu$m photometry and IRS synthesized photometry presented in Section \ref{sec:irs_obs}). However, given the different strategies in absolute flux calibration, it is not straight forward to compare photometry directly in high fidelity.

Here we complete the search for variations by focusing on the mid-infrared spectroscopy. For the HD 113766 system, as has been discussed by \citet{olofsson13}, the shape of the 10 $\mu$m feature remains similar when comparing the mid-infrared spectra obtained by VLT/VISIR in 2009 and 2012 to the 2004 {\it Spitzer} spectrum; however, the overall level is hard to quantify due to the difficulty in correcting for atmospheric absorption. Figure \ref{fig:comp_midir_spectra}a shows the comparison of the spectra taken with TIMMI2 at ESO in 2002 \citep{schutz05}, {\it Spitzer} in 2004, Subaru in 2006 and SOFIA in 2017. Overall, the shape of the 10 $\mu$m feature is very similar with a subtle (up to 15\%, $\sim$2$\sigma$) difference in the both blue- and red-side of the feature for all four spectra. In the 20 $\mu$m region, the overall flux level in the 2017 SOFIA spectrum is consistently lower than the 2004 {\it Spitzer} spectrum by $\sim$10\% (see the flux ratio in the bottom panel), but still within 1$\sigma$. Some of the large difference in flux occurs at the wavelengths where the atmospheric transmission is the worst.  We do note that the blue and red sides of the 10 $\mu$m feature cover the wavelength region where the features of sub-$\mu$m-sized silica grains are present \citep{koike13}, and the nominal atmospheric transmission in this region is above 90\%. The subtle difference in the 10 $\mu$m shape might be due to a change in the amount of silica grains, but this cannot be confirmed because of the calibration uncertainty. 

We observe a similar behavior for the mid-infrared spectra in the HD 172555 system (Figure \ref{fig:comp_midir_spectra}b). No change in the 20 $\mu$m region within 10\% is detected between {\it Spitzer} 2004 and SOFIA 2016 spectra, and the shape of the 10 $\mu$m region shows a subtle change near the wavelength region at which the silica features are present; however, the tentative change (less than 3$\sigma$) cannot be confirmed. In summary, no change (more than 20\%) is detected in the mid-infrared spectra over $\sim$15 year span for both systems. Variations of $\lesssim$10\%  might be present (particularly in the region of silica features); however, they cannot be confirmed due to calibration uncertainties across telescopes/instruments.

\section{Discussion} 
\label{sec:sec3}

\subsection{Disk Structures Inferred from SED Models and Spatially Resolved Images} 
\label{sec:diskstructure}

There are various SED models for both systems in the literature. Depending whether the fits include the prominent solid-state features and/or excess emission long-ward of 100 $\mu$m, the derived dust temperatures and the number of components in the models vary from study to study. For example, excluding the infrared bands that cover the dust features ($\sim$8--25 $\mu$m), we determine that the disk emission in HD 172555 requires only one modified blackbody ($T_d\sim$250 K) to fit the disk SED, while the one in HD 113766 requires two ($T_d\sim$490 K and 150 K) such functions\footnote{The modified black body is a function of wavelength $\lambda$ given by B$_{\nu}(\lambda, T_d)$ for $\lambda < \lambda_0$ and $\alpha(\lambda_o / \lambda)^{\beta}$B$_{\nu}(\lambda, T_d)$  for $\lambda > \lambda_0$ , where B$_{\nu}$ is the Planck function, and the temperature $T_d$ and scaling factor $\alpha$,  $\lambda_0$ and $\beta$ are free parameters. This function is widely adopted for fitting debris disk SEDs to account for the fact that the emission comes from grains in all sizes, and that grains will not act as perfect black bodies, in particular emitting inefficiently at wavelengths longer than their own size; $\lambda_0$ and $\beta$ are thus related to the grain size distribution.} (Figure \ref{fig:sed_hd172555_hd113766}). When primarily fitting the solid-state features presented in the IRS spectra, 335 K and 490 K are derived for HD 172555 and HD 113766, respectively \citep{lisse09,lisse08}. In the latter case, an additional $\sim$75 K component is required to fit the 70 $\mu$m excess, suggesting two separate components in the HD 113766 disk \citep{lisse08}. How to translate these dust temperatures to physical distance from the star  depends significantly on the underlying grain properties, particularly the minimum grain sizes in the debris \citep{thebault19}. 

High angular resolution imaging studies have also been performed for both systems from the ground, providing direct constraints on the location of the dust emission. \citet{smith12} presented a comprehensive mid-infrared study of these two systems including interferometric, spectroscopic and narrow-band imaging observations obtained with the VLT. For the HD 113766 system, \citet{olofsson13} confirmed that the disk has two distinct components by simultaneously fitting the imaging data from the VLT and the well-sampled infrared disk SED, including the solid-state features. With the pronounced dust features and the mid-infrared interferometric measurements, the inner disk in HD 113766 is relatively constrained to radii of 0.5--1.2 au (including 1$\sigma$ margin) from the star. The outer disk is estimated to lie at radii of $\sim$8--16 au (i.e., a diameter of 0\farcs5), consistent with not being resolved in the N- and Q-band images \citep{smith12}. Surprisingly, the HD 113766 system is marginally resolved at 1.24 mm by ALMA \citep{lieman-sifry16} with a synthesized beam of 1\farcs39$\times$0\farcs81, suggesting the millimeter emission from the system is extended beyond 44 au from the star. Because the bulk of the disk emission peaks at mid-infrared wavelengths (see left panel of Figure \ref{fig:sed_hd172555_hd113766}), it is difficult to reconcile the large difference in the disk sizes measured at mid-infrared and millimeter if both come from the same outer disk component. A SED model using the parameters derived by \citet{olofsson13} and a power-law extrapolation using the measured far-infrared fluxes both predict a total 1.24 mm flux that is smaller than the ALMA value by a factor of 2. This suggests that either the ALMA detection comes from a different cold component that predominantly emits in the millimeter wavelengths or that the ALMA detection is contaminated by background galaxies.

For the HD 172555 system, mid-infrared interferometric observations completely resolve out the disk emission, suggesting the bulk of debris emission is outside of 0.5 au from the star \citep{smith12}. The disk was first resolved in the thermal infrared (at $\sim$20 $\mu$m, \citealt{smith12}) and later in polarized scattered light \citep{engler18}. Mid-infrared imaging suggests that the disk is inclined by $\sim$75\arcdeg\ from face-on. Assuming a Gaussian-ring-like model, the bulk of disk emission at 20 $\mu$m is peaked at 7.7$\pm$1.5 au from the star \citep{smith12}, which is consistent with the scattered light result ($\sim$8.5--11.3 au, \citealt{engler18}). Furthermore, both studies suggest that there is significant material interior to the bulk emission of the disk (less than 8 au). The system also has an ALMA observation at 1.3 mm where both the thermal dust continuum and CO (2-1) gas emission are detected (L. Matr\`a, priv. comm.).  The disk is not spatially resolved in the dust continuum, but spectrally resolved in the CO emission that is best described by a narrow ring peaked at $\sim$8 au from the star.

In summary, the HD 113766 disk is complex and consists of an inner, $\sim$1-au component that emits mainly in the {\it Spitzer} IRAC and IRS 10--20 $\mu$m wavelengths, and an outer, $\sim$10-au component that emits its bulk emission in the far-infrared and millimeter wavelengths. On the contrary, the HD 172555 disk appears to be one broad component peaked at $\sim$8-au from the star with material existing interior and exterior to the peak radius. We note that the location of the pronounced solid-state features is near $\sim$6 au, inferred from the best-fit dust temperature \citep{lisse09}, apparently interior to the peak radius. However, given the degeneracy of SED modeling, the relative location of the material that produces these features is uncertain.

\subsection{Implications}

As has been discussed in the literature, the amount of small (sub-$\mu$m to $\mu$m size) grains that produce the prominent solid-state features must be stable over a decadal timescale, as reinforced by the new Subaru and SOFIA observations, so that the shape and relative strength of the 10 $\mu$m feature (within $\sim$10\% level) remains the same. The relative stability implies that the dust production and loss rates are balanced. The production of small grains in typical debris systems mainly comes from the collisional cascades within the swarm of particles, while the loss mechanism is dominated by collisional destruction followed by radiation pressure blowout for stars as luminous as these two.  Although the exact size of grains that are subject to radiation blowout depends sensitively on the properties of the stars (mass and luminosity) and dust grains (composition, sizes and porosity), the blowout limit is around $\sim$1 $\mu$m in both systems if one assumes compact, silicate-like composition (which generally gives the smallest blowout sizes).  \citet{arnold19} concluded that the blowout size could be an order of magnitude larger when mixing different kinds of compositions and porosity. However, the compositions used in that study are materials commonly found in solar system interplanetary particles; minor minerals such as crystalline silicates (producers of sharp solid-state features in debris disks) are not 
included. 

We first review whether the dust in these systems can be produced in typical collisional cascades. We adopted the analytical formula and criterion in \citet{wyatt07a} for the assessment. Due to the necessary assumptions and associated uncertainties, the adopted model requires a high threshold (a factor of 1000) to differentiate systems that are most likely in a transient state from the ones that are consistent with steady-state collisional cascades.  For the HD 172555 system, the following parameters were used: 1.8 $M_{\sun}$, 9 $L_{\sun}$ and 20 Myr for the star with a planetesimal belt at $r\sim$9 au. The resultant maximum infrared fractional luminosity ($f_{\rm max}$) is 2$\times$10$^{-4}$ for a typical collisional cascade system. The observed infrared fractional luminosity is only 3--4 times larger than this value, i.e., within the uncertainties the dust in the HD 172555 can be produced by a massive planetesimal belt. For the HD 113766 system, 1.5 $M_{\sun}$, 4 $L_{\sun}$ and 20 Myr were used for the star, and planetesimal belts at $r\sim$1 au and $r\sim$10 au were assumed. The corresponding $f_{\rm max}$ is 3$\times$10$^{-6}$ and 6$\times$10$^{-4}$ for the inner and outer planetesimal belts. The observed fractional luminosity is $\sim$3.3$\times$10$^{-2}$ for the inner belt, and $\sim$7$\times$10$^{-4}$ for the outer one \citep{olofsson13}. That is, the dust in the outer belt can arise in a conventional collisional cascade, but that in the inner belt is transient (i.e., the observed amount is $\sim$10,000 times the maximum value for a steady state).

If this fine dust is transient, one might expect significant changes in the solid-state features given the amount of sub-$\mu$m-sized grains ($\sim$10$^{23}$ g) and their short lifetimes ($\sim$1 yr) due to radiation blowout. However, there is no convincing case that such changes are seen.  Recent studies of two extremely dusty systems, BD+20 307 (two late F-type binary) and HD 145263 (F4V) \citep{thompson19,badlisse2020}, also suggest that there is no large-degree (more than 20\%) change in the mid-infrared spectra over a decade\footnote{See Appendix \ref{details_sofia} for the discussion about BD+20 307}. A common question in the literature has been {\it "How did such tiny grains get created in the first place?"} The concern is that it may be difficult to create such abundant sub-$\mu$m-sized grains in a typical, optically-thin, collisional-cascade debris system because of the short lifetime to radiation pressure blowout for their parent $\sim$$\mu$m-sized grains. However, \citet{thebault19} have calculated new models of collisional cascades and suggest that sub-$\mu$m grains below the blowout size can accumulate in sufficient number potentially to resolve this conundrum for bright disks around early-type stars (such as HD 172555). Their model has a number of uncertainties, such as operating only in 1-D, having to make extrapolations of the strength law toward the  sub-$\mu$m regime, etc. It would also make it puzzling that we do not observe such prominent solid-state features toward the majority of the bright disks around early-type stars with similar fractional excesses; HIP 73145 (an A2V star with $f_d\sim8\times10^{-4}$) and HIP 77315 (an A0V star with $f_d\sim2\times10^{-4}$) from \citet{ballering17} are two such examples. Furthermore, \citet{thebault19} conclude that massive debris systems around solar-like stars are not sufficiently effective at preserving grains smaller than the blowout size to result in pronounced solid-state features in the mid-infrared. Further work is needed to test their suggestion.

Although such a steady-state solution might be attractive in non-evolving systems, it is less convincing for the systems with disk variability, such as HD 113766. The recent CO gas detection (L. Matr\`a, priv. comm.) and the large amounts of atomic species \citep{riviere-marichalar14,kiefer14,grady18_hd172555} in the HD 172555 system also argue for a transient phenomenon. In these active systems, an optically-thick cloud of debris produced by a violent impact between two bodies with sizes of large asteroids or proto-planets might be a better explanation for the presence of abundant sub-$\mu$m grains. The presence of such  optically-thick debris clouds is inferred to explain the complex disk light curves observed in some extremely dusty, young debris disks \citep{meng14,su19}. Given the mass and confined volume resulting in a large impact, the impact-produced cloud is very likely to be optically thick initially before thinning by Keplerian shear. Such an environment provides perfect conditions to generate over-abundant small grains because stellar photons can only efficiently remove the small grains at the surface of the cloud. Once the impact-produced cloud becomes optically thin, the small grains that are subject to radiation pressure blowout would quickly be removed from the system, likely resulting in a rapid drop in the system's flux due to the decrease of dust temperatures, as has been observed in the ID8 system \citep{su19}. However, there may be sufficient time to generate even smaller grains within the cloud that are not subject to radiation pressure blowout, and are revealed only after the cloud becomes optically thin. This might be the case for the HD 172555 system where \citet{johnson12b} suggested that tiny, 0.02 $\mu$m-sized obsidian grains are not subject to radiation blowout. Similarly, 0.1 $\mu$m-sized silicate grains in the HD 113766 system are expected to be stable due to the same reason. In this scenario, the prominent infrared features would remain stable for a long time. The timescale on which we could expect them  to change would then be set by the (much longer) Poynting--Robertson (P--R) drag timescale of these fine grains, which is $\sim$1.3--7.7$\times10^{2}$ yr for the HD 113766 inner component and $\sim$1.6--3.6$\times10^{4}$ yr for the HD 172555 system\footnote{The P--R timescale is formulated as $\tau_{PR} = c r^2/ (4 G M_{\ast} \beta) \sim 800\ {\rm yr}\ (r/{\rm au})^2 (M_{\sun}/M_{\ast})(0.5/\beta)$ where $c$ is the speed of light, $G$ is the gravitational constant, and $M_{\ast}$ is the stellar mass. We assume $\beta$=0.5 to calculate the P--R timescale as a minimal value. }.

For the HD 113766 system, there appears to be a conflict between the variability from warm {\it Spitzer} observations and the stable solid-state features in the mid-infrared. We note that the level of variability observed at 3.6 and 4.5 $\mu$m is consistent with that of the possible changes in the mid-infrared spectra from {\it Spitzer}, Subaru and SOFIA if the dust emission is optically thin.
If some part of the 3.6/4.5 $\mu$m emission is optically thick, one would not expect the IRAC flux to track the optically-thin solid-state features in the mid-infrared.  It is also likely that the inner disk component has a background population of $\gtrsim$ km-sized planetesimals, creating additional low-level variations traced by the high-precision IRAC photometry. The tentative (2$\sigma$) positive correlation between the long-term disk flux and color temperature (see Section \ref{sec:irac_obs}) might be due to a change in dust location under the optically thin assumption. A 13\% change in the color temperature is then translated to a 27\% change in location ($T_d \sim 1/\sqrt{r}$). A dust clump/arc in a modestly eccentric ($e\sim$0.12) orbit at 1.5 au can produce such a flux/temperature variation. However, the typical grain sizes in the clump/arc have to be very small to match the observed color temperature, consistent with the hypothesis of a stable population of sub-$\mu$m grains responsible for the pronounced solid-state features. Future observations in the range of 3--5 $\mu$m can further test the hypothesis of an eccentric clump/arc in the HD 113766 inner component.

\subsection{Can Dust Composition alone Tell us How Small Grains are Generated in Young Dusty Debris Systems?}

The infrared wavelength range contains the fundamental bending, stretching and skeleton modes of solid-state species. Consequently, infrared spectroscopy of molecular clouds and protoplanetary disks \citep{vandishoeck04,sargent09} provides crucial information on the sizes and composition of dust grains, which are the building blocks embedded in $\gtrsim$ km-sized planetesimals within a planetary system. We can learn about these building blocks through infrared spectroscopy, when they are released by planetesimals shattered to form second-generation dust debris. In addition to coagulation, dust grains in the protoplanetary disks have also been through various heating/cooling processes so that crystalline forms of silicates (such as olivine, pyroxene and silica) are commonly found in the more evolved stages of planet-forming disks \citep{sargent09}, while such crystalline grains are largely absent in the interstellar medium \citep{kemper04}. Therefore, it should not be a surprise to find crystalline silicates in debris disks if these processed grains are stored in planetesimals. Nonetheless, {\it Spitzer}/IRS studies of debris systems reveal that the majority of these disks have a featureless dust continuum in the mid-infrared \citep{chen05b,mittal15}. This is understandable given that the typical blowout size is $\sim$1 $\mu$m in these systems, and that grains larger than a few $\mu$m have weak and broad features.

In contrast, conspicuous crystalline solid-state features in the mid-infrared are characteristic of extreme debris systems \citep{song05,rhee08,lisse08,weinberger11,fujiwara12,olofsson12,olofsson13,badlisse2020}, produced by warm sub-$\mu$m-sized silicates similar to the ones commonly found in Herbig Ae/Be \citep{bouwman01} and T-Tauri disks \citep{kessler06_c2d,sargent09}. Despite different modeling approaches from study to study, the strength of the features typically requires $\sim$10$^{20-25}$ g of sub-$\mu$m-sized grains in optically thin environments. These features are similar to laboratory-measured features from meteoric, terrestrial crustal, and mantle material measured in powder form, implying that the debris dust in these young systems underwent various degrees of shock and/or high temperature events \citep{morlok14,deVries18}. Although such connections in dust mineralogy between debris disks and highly altered/processed material have been found in the solar system, similar links seem to be present in the protoplanetary stage of evolution when gas was still present. Therefore, identifying the dust species alone does not tell us when these altered grains were formed either in the recent or far past. However, given the issues of retaining such a large amount of these sub-$\mu$m-sized grains (as discussed in the previous section), it becomes clear that they have to be generated in the recent past.  

Another indication of recent violent events in debris disks is the detection of freshly condensed sub-$\mu$m-sized silica smokes \citep{rhee08,lisse09,fujiwara12}. Indeed, this is thought to be the process that formed the first sub-$\mu$m-sized solids in the solar nebula, which is also verified by the laboratory experiments (see the book by \citealt{lauretta06}). However, this kind of condensation took place in a low-pressure environment, $\sim$10$^{-3}$--10$^{-6}$ bar in the early solar nebula \citep{ebel06}. Such an environment is very different from the high pressure (and temperature) condition within a vaporizing collision between two large bodies \citep{johnson12a}. The sizes of the vapor condensates generated in a violent collision between two large planetesimals depend sensitively on the collisional conditions, but are nearly all larger than 10 $\mu$m for impactors larger than 10 km \citep{johnson12a}.  Roughly mm-sized impact glasses (spherules and shards) are commonly found in the sample returns from the Apollo mission (see the review by \citealt{zellner19}). Most importantly, these impact-produced glasses will not give rise to the prominent solid-state features displayed in young dusty debris disks, unless they are broken up into much smaller grains. 

Recently, extreme space weathering has been proposed to break up larger grains and to explain the "unique" dust mineralogy in the young ($\sim$11 Myr, \citealt{pecaut12}), dusty ($f_d\sim$1.2$\times$10$^{-2}$, \citealt{mittal15}) HD 145263 system \citep{badlisse2020}. A key ingredient to the space weathering model is super massive flares or intense X-ray/UV radiation from the young star, a condition that is applicable to the AU Mic system \citep{macgregor20} and many other young late K and M-type stars. However, there is no detectable solid-state feature in the AU Mic system (see the IRS spectra provided by CASSIS project, \url{http://cassis.sirtf.com/}) and pronounced solid-state features are rarely seen in late-type disks \citep{lawler09,mittal15}. Although the energetic stellar photons might be efficiently altering the surface composition of large (a few 100 $\mu$m) grains as discussed by \citet{badlisse2020}, they most likely vaporize small $\mu$m-sized grains as suggested by \citet{osten13}. Therefore, this mechanism cannot explain the amount of sub-$\mu$m-sized grains required to produce the solid-state features observed in this system.

Using the same analytical estimate discussed in the previous section for HD 145263, we found $f_{\rm max}\sim$5.2$\times$10$^{-6}$, 2.6$\times$10$^{-5}$, and 6.7$\times$10$^{-5}$ for a planetesimal belt at $r\sim$1, 2 and 3 au assuming the same stellar parameters as in HD 113766A and an age of 11 Myr. The dust level observed in the HD 145263 system is on the borderline\footnote{We refer it as borderline because only the smallest (1-au) planetesimal belt location meets the criterion, by more than a factor of 1000, as in a transient state while others are only larger by a few 100.} between transient dust and collisional cascade dust produced in a massive planetesimal belt.  Given the amount of sub-$\mu$m-sized grains ($\sim$10$^{25}$ g, \citealt{badlisse2020}) and the short lifetime of their parent $\mu$m-sized grains in optically thin collisional cascades, we suggest that the fine dust in the HD 145263 system, similar to the two systems we discussed in this paper, was created by transient events.  A large collision, involving large-sized ($\gtrsim$100 km) asteroids that facilitated the over production of sub-$\mu$m-sized grains within an optically thick impact-produced debris clump, is a better scenario to explain the distinctive solid-state features observed in these young systems.

\section{Conclusion}
\label{sec:sec4}

We present multi-epoch, infrared photometric and spectroscopic data for two young ($\sim$20 Myr), extremely dusty debris systems around HD 113766A and HD 172555. High precision (S/N$>$100, i.e., 1\% photometric accuracy) 3.6 and 4.5 $\mu$m data were obtained during {\it Spitzer} warm mission to assess the disk variability. We found no variability for HD 172555 within 0.5\% of the average fluxes in either bands with the data taken between 2013 and 2017. These measurements are consistent with the expected photospheric values to within 1.5\% (i.e, no infrared excess). For the HD 113766 system, on the contrary, non-periodic variability was detected at $\sim$10--15\% peak-to-peak levels relative to the brightness of the primary, from 2015 to 2018. There is a 2$\sigma$ positive correlation between the long-term disk flux and color temperature, suggesting that the variation might be due to changes in dust temperature (i.e., location) if the dust emission is optically thin. 

Low-resolution, mid-infrared spectra obtained with SOFIA/FORCAST and Subaru/COMICS are presented in this study. FORCAST grism spectra covering 8.4--13.7 $\mu$m (G111 mode) and 17.6--27.7 $\mu$m (G227 mode) were obtained in 2016 and 2017 for HD 172555 and HD 113766, respectively. Using the similar grism data from bright flux standards available in the SOFIA archive, we further determined an overall 1$\sigma$ of 6\% (for G111 mode) and of 10\% (for G227) uncertainties associated with these SOFIA spectra, including the flux calibration and instrumental repeatability. Additionally, COMICS 8--13 $\mu$m spectra taken in 2006 and 2017 were presented for the HD 113766 system. We find the shape of the 10 $\mu$m feature and the overall flux level (within 10\%) are very similar between the two spectra taken 11 years apart. 

We also present multi-epoch {\it Spitzer} MIPS data obtained during the cryogenic mission including photometry at both 24 and 70 $\mu$m channels and MIPS-SED mode data. No flux variation (within 1\% at MIPS 24 $\mu$m and 5\% at MIPS 70 $\mu$m) was found for HD 172555 using data from 2004 to 2008 and HD 113766 using data from 2004 to 2005. Combined with other multi-wavelength ancillary data, these data were used to discuss the disk structure in each of the systems. The HD 113766 system has a two-component disk with an inner $\sim$1-au component that emits mainly in the $\sim$3--20 $\mu$m region, and an outer $\sim$10-au component that dominates the far-infrared and millimeter wavelengths. The HD 172555 disk appears to have one component peaked at $\sim$8 au from the star with significant amounts of material interior and exterior to the peak radius. 

We assess the temporal variability using the mid-infrared spectra obtained over a $\sim$15 year span for both systems. Using the high-quality {\it Spitzer} IRS spectrum as a reference, we found no significant (more than 20\%) change in either the shape of the prominent 10-$\mu$m solid-state feature or the overall mid-infrared flux levels for both systems. Variations of $\lesssim$10\% might be present on the blue and red sides of the 10 $\mu$m feature where the dust features from sub-$\mu$m-sized silica are generally found. However, such subtle changes cannot be confirmed due to calibration uncertainties, and need to be confirmed with future observations obtained with a more stable instrument such as JWST/MIRI.

The amount of sub-$\mu$m-sized grains required to produce the prominent 10 $\mu$m feature is on the order of 10$^{23}$ g for both systems \citep{lisse09,olofsson13}. Adopting the analytical calculation and criterion set by \citet{wyatt07a}, we verified that the amount of sub-$\mu$m-sized grains in the inner component of the HD 113766 system is unlikely to be produced in a typical collisional cascade system, while its outer component and the HD 172555 disk are consistent with steady-state collisional cascades. For massive debris disks around early-type stars (such as HD 172555), new collisional cascade calculations by \citet{thebault19} suggest that a sufficient amount of sub-$\mu$m-sized grains can accumulate and produce stable and pronounced 10 $\mu$m features. Nevertheless, some other ingredients might be missing in their model because the majority of early-type massive disks show featureless dust continua in the mid-infrared. 

Extreme space weathering has been proposed to explain the unusual dust mineralogy in the HD 145263 system \citep{badlisse2020}, where the host star is very similar to HD 113766A (both early F-type and young). We do not think space weathering is a viable explanation for the presence of abundant sub-$\mu$m-sized grains. Energetic photons from super massive flares around young solar-like stars are most likely to destroy/vaporize small dust grains present in the disk. Although intense space weathering might be able to alter the surface composition of large grains and preserve the usual composition as they broke up in subsequent collisions, this mechanism cannot explain the fact that the majority of the debris systems around young, active stars show featureless mid-infrared spectra. We also note that sub-$\mu$m-sized crystalline silicates and silica are commonly found in planet forming disks when gas was still present \citep{bouwman01,kessler06_c2d,sargent09}. Therefore, dust species alone do not inform us when these highly processed grains were formed. They could be formed in the early protoplanetary disk stage in which those altered grains were stored in planetesimals and subsequent collisional grinding among planetesimals released them into the circumstellar environment. They could also be formed in the recent past when amorphous fine grains experienced high temperature events such as violent impacts involving large asteroidal or planetary bodies. 

Finally, we suggest that disk variability might be a useful signpost to reveal highly altered grains generated in the recent past for young and dusty exoplanetary systems. In this case, such as the inner component of the HD 113766 system, the abundant sub-$\mu$m-sized grains might be due to the intense collisions among their parent mm-sized vapor condensates produced in an initially optically thick, impact-produced debris clump. Such an optically thick environment would easily lead to an over production of fine grains that are smaller than the typical blowout size in the center of the clump because stellar photons cannot easily penetrate. Once the impact-produced clump becomes optically thin as stretched by Keplerian shear, the grains subject to radiation pressure blowout (just slightly smaller than the blowout size) are quickly removed from the system, likely resulting in a rapid flux drop in the system's infrared output (a phenomenon that has been seen in the ID8 system, \citealt{su19}). If the impact-produced cloud has enough time in the optically thick condition to allow for the production of even smaller grains that are not subject to radiation blowout ($\sim$0.02 $\mu$m glassy silica in HD 172555 and $\sim$0.1 $\mu$m crystalline silicates in HD 113766A), these fine grains would be stable after the clump/arc becomes optically thin as suggested by \citet{johnson12b}. If there is no additional input for the sub-$\mu$m-size grains such as additional large impacts, the lifetime of these features should be unchanged within the P--R timescale, i.e., beyond the human lifetime for these two systems. Nevertheless, for the systems with such features located on less than a 0.2 au around 1-2 M$_{\sun}$ stars, changes within a couple of decades are possible. For the HD 113766 system, the long-term flux trend and the positive correlation between disk flux and color temperature observed by {\it Spitzer} are consistent with the presence of such an optically thin clump/arc on a modestly eccentric orbit in the inner component. The lifetime of such an impact-produced clump/arc depends sensitively on the impact condition and subsequent collisional evolution and the asymmetric phase can last for a few hundred orbital evolutions \citep{jackson14,kral15}.

\facilities{{\it Spitzer} (IRAC, IRS, MIPS), SOFIA (FORCAST), Subaru (COMICS)}

\acknowledgments

This work is based on observations made with the {\it Spitzer} Space Telescope, which is operated by the Jet Propulsion Laboratory, California Institute of Technology, and made with the NASA/DLR Stratospheric Observatory for Infrared Astronomy ({\it SOFIA}). {\it SOFIA} is jointly operated by the Universities Space Research Association, Inc. (USRA), under NASA contract NAS2-97001, and the Deutsches SOFIA Institut (DSI) under DLR contract 50 OK 0901 to the University of Stuttgart. This work has made use of data from the European Space Agency (ESA) mission {\it Gaia} (\url{https://www.cosmos.esa.int/gaia}), processed by the {\it Gaia} Data Processing and Analysis Consortium (DPAC, \url{https://www.cosmos.esa.int/web/gaia/dpac/consortium}). Funding for the DPAC
has been provided by national institutions, in particular the institutions
participating in the {\it Gaia} Multilateral Agreement. Financial support for this work was provided by NASA through award \# SOF04-0015 and SOF05-0019 issued by USRA, and ADAP program (grant No.\ NNX17AF03G). C.M.\ acknowledges support from NASA through grant 13-ADAP13-0178. Based in part on data collected at Subaru Telescope and obtained from the SMOKA, which is operated by the Astronomy Data Center, National Astronomical Observatory of Japan. We thank the anonymous referee for a thorough reading and prompt report. KYLS thanks C. Chen for providing published IRS spectra used in this work.

\appendix 
\restartappendixnumbering

\section{Warm IRAC Photometry for the HD 113766 and HD 172555 photometry}

The measured fluxes using the {\it Spitzer} IRAC observations described in Section \ref{sec:irac_obs} are given in Tables \ref{tab:irac_hd113766} and \ref{tab:irac_hd172555} for the HD 113766 and HD 172555 systems, respectively. For HD 113766, the quoted fluxes are for both the A and B components. As discussed in Section \ref{sec:irac_obs}, only the A component has an infrared excess indicative of circumstellar dust. We predicted the combined stellar output from both components in Appendix \ref{app:stellar_hd113766} and used it to derive the excess fluxes of HD 113766A at both 3.6 and 4.5 $\mu$m (also given in Table \ref{tab:irac_hd113766}). The uncertainty associated with the excess fluxes include 1.5\% of the combined stellar output. 

\begin{deluxetable*}{ccrrrrcrrrr}
\tablewidth{0pc}
\footnotesize
\tablecaption{The IRAC fluxes of the HD 113766 system\label{tab:irac_hd113766}}

\tablehead{
\colhead{AOR Key}&\colhead{BMJD$_{3.6}$}&\colhead{$F_{3.6}$}&\colhead{$E_{3.6}$}&\colhead{$exeF_{3.6}$}&\colhead{$exeE_{3.6}$}  &\colhead{BMJD$_{4.5}$}& \colhead{$F_{4.5}$}&\colhead{$eF_{4.5}$}&\colhead{$exeF_{4.5}$}&\colhead{$exeE_{4.5}$}    \\ 
\colhead{  }&\colhead{(day) }&\colhead{(mJy)}&\colhead{(mJy)}&\colhead{(mJy)}&\colhead{(mJy)}&\colhead{(day)}&\colhead{(mJy)}&\colhead{(mJy)}&\colhead{(mJy)}&\colhead{(mJy)}
}
\startdata
    53463040 &   57128.539790  &   788.99 &      3.67 &     61.09  &    11.52  &      57128.538250  &   676.37  &     1.80 &    220.82  &     7.07    \\
    53462784 &   57133.418070  &   775.56 &      4.26 &     47.66  &    11.72  &      57133.416540  &   676.96  &     1.96 &    221.41  &     7.11    \\
    53462528 &   57138.283870  &   788.66 &      3.29 &     60.76  &    11.40  &      57138.282350  &   673.29  &     2.00 &    217.74  &     7.12    \\
    53462016 &   57142.636470  &   772.22 &      4.17 &     44.32  &    11.69  &      57142.634950  &   673.03  &     1.41 &    217.49  &     6.98    \\
    53461504 &   57147.970320  &   779.14 &      4.33 &     51.24  &    11.74  &      57147.968810  &   670.18  &     1.89 &    214.63  &     7.09    \\
    53460992 &   57153.371310  &   777.25 &      4.44 &     49.35  &    11.79  &      57153.369770  &   671.43  &     1.60 &    215.88  &     7.02    \\
\enddata
\tablecomments{$F$ and $E$ are the flux and uncertainty including the
star, while $exeF$ and $exeE$ are the excess quantities excluding the
star.  This table is published in its entirety in the
machine-readable format. A portion is shown here for guidance regarding its form and content.}
\end{deluxetable*}

\begin{deluxetable*}{ccrrcrr}
\tablewidth{0pc}
\footnotesize
\tablecaption{The IRAC fluxes of the HD 172555 system\label{tab:irac_hd172555}}

\tablehead{
\colhead{AOR Key}&\colhead{BMJD$_{3.6}$}&\colhead{$F_{3.6}$}&\colhead{$E_{3.6}$}&\colhead{BMJD$_{4.5}$}& \colhead{$F_{4.5}$}&\colhead{$eF_{4.5}$}   \\ 
\colhead{  }&\colhead{(day) }&\colhead{(mJy)}&\colhead{(mJy)}&\colhead{(day)}&\colhead{(mJy)}&\colhead{(mJy)}
}
\startdata
  48318464 &    56455.24044 &  5316.18 &     0.85 &    \nodata     &  \nodata &  \nodata \\
  58766592 &    57561.96057 &  5358.98 &     1.35 &    57561.96115 &  3501.75 &     1.03  \\ 
  58766336 &    57571.61957 &  5352.53 &     0.75 &    57571.62014 &  3495.64 &     1.30  \\ 
  58766080 &    57581.74171 &  5366.15 &     1.51 &    57581.74228 &  3504.02 &     1.19  \\ 
  58765824 &    57591.92158 &  5337.11 &     0.85 &    57591.92215 &  3497.69 &     0.87  \\ 
  58765568 &    57601.00141 &  5335.97 &     0.79 &    57601.00198 &  3491.22 &     1.15  \\ 
  58765312 &    57716.17711 &  5325.11 &     1.22 &    57716.17770 &  3484.07 &     1.03  \\ 
  58765056 &    57725.60858 &  5334.20 &     0.74 &    57725.60915 &  3493.79 &     1.48  \\ 
  58764800 &    57736.09562 &  5384.17 &     1.45 &    57736.09620 &  3486.51 &     1.28  \\ 
  58764544 &    57747.12971 &  5326.71 &     0.65 &    57747.13028 &  3498.77 &     0.94  \\ 
  58764288 &    57755.27977 &  5337.39 &     1.00 &    57755.28034 &  3510.49 &     1.08  \\ 
\enddata
\tablecomments{The average flux densities are 5340$\pm$20 mJy and 3496 $\pm$8 mJy at the 3.6 and 4.5 $\mu$m bands, respectively. }
\end{deluxetable*}

\section{Stellar Properties for the HD 113766 System}
\label{app:stellar_hd113766}

We used the photometry that contains both stars (Johnson UVB and 2MASS JHK) to estimate the combined stellar output in the IRAC wavelengths. A Kurucz model with a temperature of 6750 K provides a good fit to the combined photometry, and results a total luminosity of 9.4 $L_{\sun}$ at a distance of 111 pc \citep{gaia16,gaia18}. The derived combined luminosity is within $\sim$10\% of the combined value derived by \citet{pecaut12} after correcting for the adopted distance. The expected photospheric flux is 728 mJy at IRAC 3.6 $\mu$m band, and 456 at IRAC 4.5 $\mu$m band.

\section{Details about the SOFIA/FORCAST Observations and Associated Uncertainty}
\label{details_sofia}

As described in Section \ref{sec:sofia_obs}, we used the Level-3 data products provided by the SOFIA Science Center
for further analysis. Table \ref{tab:sofia_obs} gives the details about the SOFIA/FORCAST observations presented in this work, including the time of the observation, flight altitude, instrumental grism/filter, and total integration time. 
To assess the flux calibration and repeatability in the FORCAST data, we also made used of multiyear, archival data for
bright calibrators: $\alpha$ Tau, $\alpha$ Boo, $\beta$ And, and $\sigma$ Lib. We used the imaging data to determine 
the appropriate aperture correction factor. A total of 206 "calibrated" Level-3 images was used for the F111 filter with data taken from 2013 to 2019, and a total of 22 images taken in 2016-2019 was used for the F242 filter. The aperture correction factor
for a specific setting in the aperture was determined by referencing the total flux derived using the nominal aperture setting [12,15,25] (i.e.,  a radius of 12 pixels and sky annulus of 15--25 pixels). For an aperture setting of [8,8,12], the
aperture correction factor is 1.073$\pm$0.010 for the F111 filter, and 1.121$\pm$0.042 for the F242 filter. For an aperture setting of [6,6,10], the factors are 1.127$\pm$0.018 and 1.204$\pm$0.058 for the F111 and F242 filters, respectively. Using the nominal aperture, we also confirmed the 6\% flux calibration in the photometry using these bright calibrators. 
 
For the grism data, only $\alpha$ Tau has more than a dozen G111 Level-3 ("combspec") data points taken in 2016--2018, $\sigma$ Lib has a handful of G227 data taken in 2014--2015, and the other two stars have a few spectra available in the archive. Note that these calibration data were taken at different flights, altitudes and slit widths (2\farcs4 and 4\farcs7); so the comparison among them reflects the overall repeatability in the long-term (multiyear) flux calibration, which is essential
in assessing source variability using the grism data. For a target with spectra taken in the same flight but with two slit widths, the spectra agree within a few \%. For the G111 mode, an average of 4\% repeatability is derived, but could be up to 10\% level in the ozone band. For the G227 mode, an average of 10\% repeatability is derived, but could be up to 20\%. Including the 6\% flux calibration error derived from photometry, the grism spectra have a long-term repeatability of 6\% and 10\% for the G111 and G227 mode, respectively. In summary, flux variation at the levels of 6\% and 10\% (1 $\sigma$) for the G111 and G227 grism spectra is not significant. Recently, a flux increased ($\sim$10$\pm$2\%) in the BD+20 307 system was reported by \citet{thompson19} using the 2006 {\it Spitzer} and 2015 SOFIA spectrum (G111 mode). Taking the uncertainties in absolute flux calibration, stability of the instruments, and slit loss and aperture corrections, the reported uncertainty is slightly under estimated, suggesting the tentative flux increase is at $\sim$2$\sigma$ levels.

\begin{deluxetable*}{cccclc}
\tablewidth{0pc}
\tablecaption{SOFIA Observations\label{tab:sofia_obs}}
\tablehead{
\colhead{Level-3 \#} & \colhead{Filter/Grism} & \colhead{Date} & \colhead{UT} & \colhead{Altitude (ft)} & \colhead{Integration (s)}  
}
\startdata 
\multicolumn{6}{c}{\underline{HD 172555}} \\ 
0191   &   F111 & 2016 Jul 20 &  14:38:40.3  &  43019--43020  & 26.7  \\ 
0192-0193 & F242 & 2016 Jul 20 & 14:39:45.4  &  43015--43021  & 62.6  \\
0194   &   F111 & 2016 Jul 20 &  14:42:59.8  &  43006         & 55.3  \\ 
0195   &   F111 & 2016 Jul 20 &  14:44:06.6  &  43006--43011  & 55.3  \\ 
0196   &   F111 & 2016 Jul 20 &  14:45:32.7  &  43010--43013  & 55.3  \\ 
0197   &   F111 & 2016 Jul 20 &  14:46:39.3  &  43008--43010  & 55.3  \\ 
0198$^{a}$   &   F111 & 2016 Jul 20 &  14:50:58.4  &  43017--43018  & 55.3  \\
0199   &   F111 & 2016 Jul 20 &  14:53:04.3  &  43017--43018  & 55.3  \\ 
0200   &   F111 & 2016 Jul 20 &  14:54:11.2  &  43018--43019  & 55.3  \\ 
0201-0338 & G227 & 2016 Jul 20 & 14:55:54.3  &  43017--43031  & 4496.7 \\
0287-0329 & G111 & 2016 Jul 20 & 16:23:39.7  &  43017--43031  & 2010.5 \\ 
\multicolumn{6}{c}{} \\ 
\multicolumn{6}{c}{\underline{HD 113766}} \\ 
0084-0089 &   F111 & 2017 Aug 03 &  09:52:55.1  &  38015--38022  & 167.2  \\ 
0090-0091 &   F242 & 2017 Aug 03 &  09:58:33.9  &  38010--38020  & 44.9  \\
0092      &   F111 & 2017 Aug 03 &  10:00:31.0  &  38014--38019  & 18.6  \\ 
0193      &   F111 & 2017 Aug 03 &  10:00:57.4  &  38015--38016  & 18.6  \\ 
0096-0127 &   G111 & 2017 Aug 03 &  10:02:54.7  &  38013--39002  & 1560.5 \\ 
0010      &   F111 & 2017 Aug 06 &  09:16:30.4  &  37085--37079  & 18.9  \\ 
0011      &   F111 & 2017 Aug 06 &  09:16:56.4  &  37068--37079  & 18.9  \\ 
0014-0077 &   G227 & 2017 Aug 06 &  09:19:02.3  &  37110--38095  & 3098.8 
\enddata 
\tablenotetext{a}{No source was found; not used for further analysis.}
\end{deluxetable*}


\begin{thebibliography}{}
\expandafter\ifx\csname natexlab\endcsname\relax\def\natexlab#1{#1}\fi

\bibitem[{{Arnold} {et~al.}(2019){Arnold}, {Weinberger}, {Videen}, \&
  {Zubko}}]{arnold19}
{Arnold}, J.~A., {Weinberger}, A.~J., {Videen}, G., \& {Zubko}, E.~S. 2019,
  \aj, 157, 157

\bibitem[{{Baba} {et~al.}(2002){Baba}, {Yasuda}, {Ichikawa}, {Yagi}, {Iwamoto},
  {Takata}, {Horaguchi}, {Taga}, {Watanabe}, {Okumura}, {Ozawa}, {Yamamoto}, \&
  {Hamabe}}]{baba02}
{Baba}, H., {Yasuda}, N., {Ichikawa}, S.-I., {et~al.} 2002, Report of the
  National Astronomical Observatory of Japan, 6, 23

\bibitem[{{Ballering} {et~al.}(2017){Ballering}, {Rieke}, {Su}, \&
  {G{\'a}sp{\'a}r}}]{ballering17}
{Ballering}, N.~P., {Rieke}, G.~H., {Su}, K. Y.~L., \& {G{\'a}sp{\'a}r}, A.
  2017, \apj, 845, 120

\bibitem[{{Balog} {et~al.}(2009){Balog}, {Kiss}, {Vink{\'o}}, {Rieke},
  {Muzerolle}, {G{\'a}sp{\'a}r}, {Young}, \& {Gorlova}}]{balog09}
{Balog}, Z., {Kiss}, L.~L., {Vink{\'o}}, J., {et~al.} 2009, ApJ, 698, 1989

\bibitem[{{Bohlin} {et~al.}(2011){Bohlin}, {Gordon}, {Rieke}, {Ardila},
  {Carey}, {Deustua}, {Engelbracht}, {Ferguson}, {Flanagan}, {Kalirai},
  {Meixner}, {Noriega-Crespo}, {Su}, \& {Tremblay}}]{bohlin11}
{Bohlin}, R.~C., {Gordon}, K.~D., {Rieke}, G.~H., {et~al.} 2011, \aj, 141, 173

\bibitem[{{Bouwman} {et~al.}(2001){Bouwman}, {Meeus}, {de Koter}, {Hony},
  {Dominik}, \& {Waters}}]{bouwman01}
{Bouwman}, J., {Meeus}, G., {de Koter}, A., {et~al.} 2001, \aap, 375, 950

\bibitem[{{Chen} {et~al.}(2011){Chen}, {Mamajek}, {Bitner}, {Pecaut}, {Su}, \&
  {Weinberger}}]{chen11}
{Chen}, C.~H., {Mamajek}, E.~E., {Bitner}, M.~A., {et~al.} 2011, ApJ, 738, 122

\bibitem[{{Chen} {et~al.}(2020){Chen}, {Su}, \& {Xu}}]{chen_su_xu20}
{Chen}, C.~H., {Su}, K. Y.~L., \& {Xu}, S. 2020, Nature Astronomy, 4, 328

\bibitem[{{Chen} {et~al.}(2005){Chen}, {Patten}, {Werner}, {Dowell},
  {Stapelfeldt}, {Song}, {Stauffer}, {Blaylock}, {Gordon}, \&
  {Krause}}]{chen05b}
{Chen}, C.~H., {Patten}, B.~M., {Werner}, M.~W., {et~al.} 2005, ApJ, 634, 1372

\bibitem[{{Chen} {et~al.}(2006){Chen}, {Sargent}, {Bohac}, {Kim},
  {Leibensperger}, {Jura}, {Najita}, {Forrest}, {Watson}, {Sloan}, \&
  {Keller}}]{chen06}
{Chen}, C.~H., {Sargent}, B.~A., {Bohac}, C., {et~al.} 2006, ApJs, 166, 351

\bibitem[{{Cutri} \& {et al.}(2014)}]{cutri14}
{Cutri}, R.~M., \& {et al.} 2014, VizieR Online Data Catalog, II/328

\bibitem[{{de Vries} {et~al.}(2018){de Vries}, {Skogby}, {Waters}, \&
  {Min}}]{deVries18}
{de Vries}, B.~L., {Skogby}, H., {Waters}, L.~B.~F.~M., \& {Min}, M. 2018,
  \icarus, 307, 400

\bibitem[{{Ebel}(2006)}]{ebel06}
{Ebel}, D.~S. 2006, {Condensation of Rocky Material in Astrophysical
  Environments}, ed. D.~S. {Lauretta} \& H.~Y. {McSween}, 253

\bibitem[{{Engler} {et~al.}(2018){Engler}, {Schmid}, {Quanz}, {Avenhaus}, \&
  {Bazzon}}]{engler18}
{Engler}, N., {Schmid}, H.~M., {Quanz}, S.~P., {Avenhaus}, H., \& {Bazzon}, A.
  2018, \aap, 618, A151

\bibitem[{{Fabricius} \& {Makarov}(2000)}]{fabricius00}
{Fabricius}, C., \& {Makarov}, V.~V. 2000, \aap, 356, 141

\bibitem[{{Fujiwara} {et~al.}(2012){Fujiwara}, {Onaka}, {Yamashita},
  {Ishihara}, {Kataza}, {Fukagawa}, {Takeda}, \& {Murakami}}]{fujiwara12}
{Fujiwara}, H., {Onaka}, T., {Yamashita}, T., {et~al.} 2012, ApJL, 749, L29

\bibitem[{{Gaia Collaboration} {et~al.}(2016){Gaia Collaboration}, {Prusti},
  {de Bruijne}, {Brown}, {Vallenari}, {Babusiaux}, {Bailer-Jones}, {Bastian},
  {Biermann}, {Evans}, {Eyer}, {Jansen}, {Jordi}, {Klioner}, {Lammers},
  {Lindegren}, {Luri}, {Mignard}, {Milligan}, {Panem}, {Poinsignon},
  {Pourbaix}, {Randich}, {Sarri}, {Sartoretti}, {Siddiqui}, {Soubiran},
  {Valette}, {van Leeuwen}, {Walton}, {Aerts}, {Arenou}, {Cropper}, {Drimmel},
  {H{\o}g}, {Katz}, {Lattanzi}, {O'Mullane}, {Grebel}, {Holland}, {Huc},
  {Passot}, {Bramante}, {Cacciari}, {Casta{\~n}eda}, {Chaoul}, {Cheek}, {De
  Angeli}, {Fabricius}, {Guerra}, {Hern{\'a}ndez}, {Jean-Antoine-Piccolo},
  {Masana}, {Messineo}, {Mowlavi}, {Nienartowicz}, {Ord{\'o}{\~n}ez-Blanco},
  {Panuzzo}, {Portell}, {Richards}, {Riello}, {Seabroke}, {Tanga},
  {Th{\'e}venin}, {Torra}, {Els}, {Gracia-Abril}, {Comoretto},
  {Garcia-Reinaldos}, {Lock}, {Mercier}, {Altmann}, {Andrae}, {Astraatmadja},
  {Bellas-Velidis}, {Benson}, {Berthier}, {Blomme}, {Busso}, {Carry},
  {Cellino}, {Clementini}, {Cowell}, {Creevey}, {Cuypers}, {Davidson}, {De
  Ridder}, {de Torres}, {Delchambre}, {Dell'Oro}, {Ducourant}, {Fr{\'e}mat},
  {Garc{\'\i}a-Torres}, {Gosset}, {Halbwachs}, {Hambly}, {Harrison}, {Hauser},
  {Hestroffer}, {Hodgkin}, {Huckle}, {Hutton}, {Jasniewicz}, {Jordan},
  {Kontizas}, {Korn}, {Lanzafame}, {Manteiga}, {Moitinho}, {Muinonen},
  {Osinde}, {Pancino}, {Pauwels}, {Petit}, {Recio-Blanco}, {Robin}, {Sarro},
  {Siopis}, {Smith}, {Smith}, {Sozzetti}, {Thuillot}, {van Reeven}, {Viala},
  {Abbas}, {Abreu Aramburu}, {Accart}, {Aguado}, {Allan}, {Allasia},
  {Altavilla}, {{\'A}lvarez}, {Alves}, {Anderson}, {Andrei}, {Anglada Varela},
  {Antiche}, {Antoja}, {Ant{\'o}n}, {Arcay}, {Atzei}, {Ayache}, {Bach},
  {Baker}, {Balaguer-N{\'u}{\~n}ez}, {Barache}, {Barata}, {Barbier}, {Barblan},
  {Baroni}, {Barrado y Navascu{\'e}s}, {Barros}, {Barstow}, {Becciani},
  {Bellazzini}, {Bellei}, {Bello Garc{\'\i}a}, {Belokurov}, {Bendjoya},
  {Berihuete}, {Bianchi}, {Bienaym{\'e}}, {Billebaud}, {Blagorodnova},
  {Blanco-Cuaresma}, {Boch}, {Bombrun}, {Borrachero}, {Bouquillon}, {Bourda},
  {Bouy}, {Bragaglia}, {Breddels}, {Brouillet}, {Br{\"u}semeister},
  {Bucciarelli}, {Budnik}, {Burgess}, {Burgon}, {Burlacu}, {Busonero}, {Buzzi},
  {Caffau}, {Cambras}, {Campbell}, {Cancelliere}, {Cantat-Gaudin}, {Carlucci},
  {Carrasco}, {Castellani}, {Charlot}, {Charnas}, {Charvet}, {Chassat},
  {Chiavassa}, {Clotet}, {Cocozza}, {Collins}, {Collins}, {Costigan}, {Crifo},
  {Cross}, {Crosta}, {Crowley}, {Dafonte}, {Damerdji}, {Dapergolas}, {David},
  {David}, {De Cat}, {de Felice}, {de Laverny}, {De Luise}, {De March}, {de
  Martino}, {de Souza}, {Debosscher}, {del Pozo}, {Delbo}, {Delgado},
  {Delgado}, {di Marco}, {Di Matteo}, {Diakite}, {Distefano}, {Dolding}, {Dos
  Anjos}, {Drazinos}, {Dur{\'a}n}, {Dzigan}, {Ecale}, {Edvardsson}, {Enke},
  {Erdmann}, {Escolar}, {Espina}, {Evans}, {Eynard Bontemps}, {Fabre},
  {Fabrizio}, {Faigler}, {Falc{\~a}o}, {Farr{\`a}s Casas}, {Faye}, {Federici},
  {Fedorets}, {Fern{\'a}ndez-Hern{\'a}ndez}, {Fernique}, {Fienga}, {Figueras},
  {Filippi}, {Findeisen}, {Fonti}, {Fouesneau}, {Fraile}, {Fraser}, {Fuchs},
  {Furnell}, {Gai}, {Galleti}, {Galluccio}, {Garabato}, {Garc{\'\i}a-Sedano},
  {Gar{\'e}}, {Garofalo}, {Garralda}, {Gavras}, {Gerssen}, {Geyer}, {Gilmore},
  {Girona}, {Giuffrida}, {Gomes}, {Gonz{\'a}lez-Marcos},
  {Gonz{\'a}lez-N{\'u}{\~n}ez}, {Gonz{\'a}lez-Vidal}, {Granvik}, {Guerrier},
  {Guillout}, {Guiraud}, {G{\'u}rpide}, {Guti{\'e}rrez-S{\'a}nchez}, {Guy},
  {Haigron}, {Hatzidimitriou}, {Haywood}, {Heiter}, {Helmi}, {Hobbs},
  {Hofmann}, {Holl}, {Holland }, {Hunt}, {Hypki}, {Icardi}, {Irwin}, {Jevardat
  de Fombelle}, {Jofr{\'e}}, {Jonker}, {Jorissen}, {Julbe}, {Karampelas},
  {Kochoska}, {Kohley}, {Kolenberg}, {Kontizas}, {Koposov}, {Kordopatis},
  {Koubsky}, {Kowalczyk}, {Krone-Martins}, {Kudryashova}, {Kull}, {Bachchan},
  {Lacoste-Seris}, {Lanza}, {Lavigne}, {Le Poncin-Lafitte}, {Lebreton},
  {Lebzelter}, {Leccia}, {Leclerc}, {Lecoeur-Taibi}, {Lemaitre}, {Lenhardt},
  {Leroux}, {Liao}, {Licata}, {Lindstr{\o}m}, {Lister}, {Livanou}, {Lobel},
  {L{\"o}ffler}, {L{\'o}pez}, {Lopez-Lozano}, {Lorenz}, {Loureiro},
  {MacDonald}, {Magalh{\~a}es Fernandes}, {Managau}, {Mann}, {Mantelet},
  {Marchal}, {Marchant}, {Marconi}, {Marie}, {Marinoni}, {Marrese},
  {Marschalk{\'o}}, {Marshall}, {Mart{\'\i}n-Fleitas}, {Martino}, {Mary},
  {Matijevi{\v{c}}}, {Mazeh}, {McMillan}, {Messina}, {Mestre}, {Michalik},
  {Millar}, {Miranda}, {Molina}, {Molinaro}, {Molinaro}, {Moln{\'a}r},
  {Moniez}, {Montegriffo}, {Monteiro}, {Mor}, {Mora}, {Morbidelli}, {Morel},
  {Morgenthaler}, {Morley}, {Morris}, {Mulone}, {Muraveva}, {Musella},
  {Narbonne}, {Nelemans}, {Nicastro}, {Noval}, {Ord{\'e}novic},
  {Ordieres-Mer{\'e}}, {Osborne}, {Pagani}, {Pagano}, {Pailler}, {Palacin},
  {Palaversa}, {Parsons}, {Paulsen}, {Pecoraro}, {Pedrosa}, {Pentik{\"a}inen},
  {Pereira}, {Pichon}, {Piersimoni}, {Pineau}, {Plachy}, {Plum}, {Poujoulet},
  {Pr{\v{s}}a}, {Pulone}, {Ragaini}, {Rago}, {Rambaux}, {Ramos-Lerate},
  {Ranalli}, {Rauw}, {Read}, {Regibo}, {Renk}, {Reyl{\'e}}, {Ribeiro},
  {Rimoldini}, {Ripepi}, {Riva}, {Rixon}, {Roelens}, {Romero-G{\'o}mez},
  {Rowell}, {Royer}, {Rudolph}, {Ruiz-Dern}, {Sadowski}, {Sagrist{\`a}
  Sell{\'e}s}, {Sahlmann}, {Salgado}, {Salguero}, {Sarasso}, {Savietto},
  {Schnorhk}, {Schultheis}, {Sciacca}, {Segol}, {Segovia}, {Segransan},
  {Serpell}, {Shih}, {Smareglia}, {Smart}, {Smith}, {Solano}, {Solitro},
  {Sordo}, {Soria Nieto}, {Souchay}, {Spagna}, {Spoto}, {Stampa}, {Steele},
  {Steidelm{\"u}ller}, {Stephenson}, {Stoev}, {Suess}, {S{\"u}veges}, {Surdej},
  {Szabados}, {Szegedi-Elek}, {Tapiador}, {Taris}, {Tauran}, {Taylor},
  {Teixeira}, {Terrett}, {Tingley}, {Trager}, {Turon}, {Ulla}, {Utrilla},
  {Valentini}, {van Elteren}, {Van Hemelryck}, {van Leeuwen}, {Varadi},
  {Vecchiato}, {Veljanoski}, {Via}, {Vicente}, {Vogt}, {Voss}, {Votruba},
  {Voutsinas}, {Walmsley}, {Weiler}, {Weingrill}, {Werner}, {Wevers},
  {Whitehead}, {Wyrzykowski}, {Yoldas}, {{\v{Z}}erjal}, {Zucker}, {Zurbach},
  {Zwitter}, {Alecu}, {Allen}, {Allende Prieto}, {Amorim},
  {Anglada-Escud{\'e}}, {Arsenijevic}, {Azaz}, {Balm}, {Beck}, {Bernstein},
  {Bigot}, {Bijaoui}, {Blasco}, {Bonfigli}, {Bono}, {Boudreault}, {Bressan},
  {Brown}, {Brunet}, {Bunclark}, {Buonanno}, {Butkevich}, {Carret}, {Carrion},
  {Chemin}, {Ch{\'e}reau}, {Corcione}, {Darmigny}, {de Boer}, {de Teodoro}, {de
  Zeeuw}, {Delle Luche}, {Domingues}, {Dubath}, {Fodor}, {Fr{\'e}zouls},
  {Fries}, {Fustes}, {Fyfe}, {Gallardo}, {Gallegos}, {Gardiol}, {Gebran},
  {Gomboc}, {G{\'o}mez}, {Grux}, {Gueguen}, {Heyrovsky}, {Hoar}, {Iannicola},
  {Isasi Parache}, {Janotto}, {Joliet}, {Jonckheere}, {Keil}, {Kim},
  {Klagyivik}, {Klar}, {Knude}, {Kochukhov}, {Kolka}, {Kos}, {Kutka}, {Lainey},
  {LeBouquin}, {Liu}, {Loreggia}, {Makarov}, {Marseille}, {Martayan},
  {Martinez-Rubi}, {Massart}, {Meynadier}, {Mignot}, {Munari}, {Nguyen},
  {Nordlander}, {Ocvirk}, {O'Flaherty}, {Olias Sanz}, {Ortiz}, {Osorio},
  {Oszkiewicz}, {Ouzounis}, {Palmer}, {Park}, {Pasquato}, {Peltzer}, {Peralta},
  {P{\'e}turaud}, {Pieniluoma}, {Pigozzi}, {Poels}, {Prat}, {Prod'homme},
  {Raison}, {Rebordao}, {Risquez}, {Rocca-Volmerange}, {Rosen}, {Ruiz-Fuertes},
  {Russo}, {Sembay}, {Serraller Vizcaino}, {Short}, {Siebert}, {Silva},
  {Sinachopoulos}, {Slezak}, {Soffel}, {Sosnowska}, {Strai{\v{z}}ys}, {ter
  Linden}, {Terrell}, {Theil}, {Tiede}, {Troisi}, {Tsalmantza}, {Tur},
  {Vaccari}, {Vachier}, {Valles}, {Van Hamme}, {Veltz}, {Virtanen}, {Wallut},
  {Wichmann}, {Wilkinson}, {Ziaeepour}, \& {Zschocke}}]{gaia16}
{Gaia Collaboration}, {Prusti}, T., {de Bruijne}, J.~H.~J., {et~al.} 2016,
  \aap, 595, A1

\bibitem[{{Gaia Collaboration} {et~al.}(2018){Gaia Collaboration}, {Brown},
  {Vallenari}, {Prusti}, {de Bruijne}, {Babusiaux}, {Bailer-Jones}, {Biermann},
  {Evans}, {Eyer}, {Jansen}, {Jordi}, {Klioner}, {Lammers}, {Lindegren},
  {Luri}, {Mignard}, {Panem}, {Pourbaix}, {Randich}, {Sartoretti}, {Siddiqui},
  {Soubiran}, {van Leeuwen}, {Walton}, {Arenou}, {Bastian}, {Cropper},
  {Drimmel}, {Katz}, {Lattanzi}, {Bakker}, {Cacciari}, {Casta{\~n}eda},
  {Chaoul}, {Cheek}, {De Angeli}, {Fabricius}, {Guerra}, {Holl}, {Masana},
  {Messineo}, {Mowlavi}, {Nienartowicz}, {Panuzzo}, {Portell}, {Riello},
  {Seabroke}, {Tanga}, {Th{\'e}venin}, {Gracia-Abril}, {Comoretto},
  {Garcia-Reinaldos}, {Teyssier}, {Altmann}, {Andrae}, {Audard},
  {Bellas-Velidis}, {Benson}, {Berthier}, {Blomme}, {Burgess}, {Busso},
  {Carry}, {Cellino}, {Clementini}, {Clotet}, {Creevey}, {Davidson}, {De
  Ridder}, {Delchambre}, {Dell'Oro}, {Ducourant},
  {Fern{\'a}ndez-Hern{\'a}ndez}, {Fouesneau}, {Fr{\'e}mat}, {Galluccio},
  {Garc{\'\i}a-Torres}, {Gonz{\'a}lez-N{\'u}{\~n}ez}, {Gonz{\'a}lez-Vidal},
  {Gosset}, {Guy}, {Halbwachs}, {Hambly}, {Harrison}, {Hern{\'a}ndez},
  {Hestroffer}, {Hodgkin}, {Hutton}, {Jasniewicz}, {Jean-Antoine-Piccolo},
  {Jordan}, {Korn}, {Krone-Martins}, {Lanzafame}, {Lebzelter}, {L{\"o}ffler},
  {Manteiga}, {Marrese}, {Mart{\'\i}n-Fleitas}, {Moitinho}, {Mora}, {Muinonen},
  {Osinde}, {Pancino}, {Pauwels}, {Petit}, {Recio-Blanco}, {Richards},
  {Rimoldini}, {Robin}, {Sarro}, {Siopis}, {Smith}, {Sozzetti}, {S{\"u}veges},
  {Torra}, {van Reeven}, {Abbas}, {Abreu Aramburu}, {Accart}, {Aerts},
  {Altavilla}, {{\'A}lvarez}, {Alvarez}, {Alves}, {Anderson}, {Andrei},
  {Anglada Varela}, {Antiche}, {Antoja}, {Arcay}, {Astraatmadja}, {Bach},
  {Baker}, {Balaguer-N{\'u}{\~n}ez}, {Balm}, {Barache}, {Barata}, {Barbato},
  {Barblan}, {Barklem}, {Barrado}, {Barros}, {Barstow}, {Bartholom{\'e}
  Mu{\~n}oz}, {Bassilana}, {Becciani}, {Bellazzini}, {Berihuete}, {Bertone},
  {Bianchi}, {Bienaym{\'e}}, {Blanco-Cuaresma}, {Boch}, {Boeche}, {Bombrun},
  {Borrachero}, {Bossini}, {Bouquillon}, {Bourda}, {Bragaglia}, {Bramante},
  {Breddels}, {Bressan}, {Brouillet}, {Br{\"u}semeister}, {Brugaletta},
  {Bucciarelli}, {Burlacu}, {Busonero}, {Butkevich}, {Buzzi}, {Caffau},
  {Cancelliere}, {Cannizzaro}, {Cantat-Gaudin}, {Carballo}, {Carlucci},
  {Carrasco}, {Casamiquela}, {Castellani}, {Castro-Ginard}, {Charlot},
  {Chemin}, {Chiavassa}, {Cocozza}, {Costigan}, {Cowell}, {Crifo}, {Crosta},
  {Crowley}, {Cuypers}, {Dafonte}, {Damerdji}, {Dapergolas}, {David}, {David},
  {de Laverny}, {De Luise}, {De March}, {de Martino}, {de Souza}, {de Torres},
  {Debosscher}, {del Pozo}, {Delbo}, {Delgado}, {Delgado}, {Di Matteo},
  {Diakite}, {Diener}, {Distefano}, {Dolding}, {Drazinos}, {Dur{\'a}n},
  {Edvardsson}, {Enke}, {Eriksson}, {Esquej}, {Eynard Bontemps}, {Fabre},
  {Fabrizio}, {Faigler}, {Falc{\~a}o}, {Farr{\`a}s Casas}, {Federici},
  {Fedorets}, {Fernique}, {Figueras}, {Filippi}, {Findeisen}, {Fonti},
  {Fraile}, {Fraser}, {Fr{\'e}zouls}, {Gai}, {Galleti}, {Garabato},
  {Garc{\'\i}a-Sedano}, {Garofalo}, {Garralda}, {Gavel}, {Gavras}, {Gerssen},
  {Geyer}, {Giacobbe}, {Gilmore}, {Girona}, {Giuffrida}, {Glass}, {Gomes},
  {Granvik}, {Gueguen}, {Guerrier}, {Guiraud}, {Guti{\'e}rrez-S{\'a}nchez},
  {Haigron}, {Hatzidimitriou}, {Hauser}, {Haywood}, {Heiter}, {Helmi}, {Heu},
  {Hilger}, {Hobbs}, {Hofmann}, {Holland}, {Huckle}, {Hypki}, {Icardi},
  {Jan{\ss}en}, {Jevardat de Fombelle}, {Jonker}, {Juh{\'a}sz}, {Julbe},
  {Karampelas}, {Kewley}, {Klar}, {Kochoska}, {Kohley}, {Kolenberg},
  {Kontizas}, {Kontizas}, {Koposov}, {Kordopatis}, {Kostrzewa-Rutkowska},
  {Koubsky}, {Lambert}, {Lanza}, {Lasne}, {Lavigne}, {Le Fustec}, {Le
  Poncin-Lafitte}, {Lebreton}, {Leccia}, {Leclerc}, {Lecoeur-Taibi},
  {Lenhardt}, {Leroux}, {Liao}, {Licata}, {Lindstr{\o}m}, {Lister}, {Livanou},
  {Lobel}, {L{\'o}pez}, {Managau}, {Mann}, {Mantelet}, {Marchal}, {Marchant},
  {Marconi}, {Marinoni}, {Marschalk{\'o}}, {Marshall}, {Martino}, {Marton},
  {Mary}, {Massari}, {Matijevi{\v{c}}}, {Mazeh}, {McMillan}, {Messina},
  {Michalik}, {Millar}, {Molina}, {Molinaro}, {Moln{\'a}r}, {Montegriffo},
  {Mor}, {Morbidelli}, {Morel}, {Morris}, {Mulone}, {Muraveva}, {Musella},
  {Nelemans}, {Nicastro}, {Noval}, {O'Mullane}, {Ord{\'e}novic},
  {Ord{\'o}{\~n}ez-Blanco}, {Osborne}, {Pagani}, {Pagano}, {Pailler},
  {Palacin}, {Palaversa}, {Panahi}, {Pawlak}, {Piersimoni}, {Pineau}, {Plachy},
  {Plum}, {Poggio}, {Poujoulet}, {Pr{\v{s}}a}, {Pulone}, {Racero}, {Ragaini},
  {Rambaux}, {Ramos-Lerate}, {Regibo}, {Reyl{\'e}}, {Riclet}, {Ripepi}, {Riva},
  {Rivard}, {Rixon}, {Roegiers}, {Roelens}, {Romero-G{\'o}mez}, {Rowell},
  {Royer}, {Ruiz-Dern}, {Sadowski}, {Sagrist{\`a} Sell{\'e}s}, {Sahlmann},
  {Salgado}, {Salguero}, {Sanna}, {Santana-Ros}, {Sarasso}, {Savietto},
  {Schultheis}, {Sciacca}, {Segol}, {Segovia}, {S{\'e}gransan}, {Shih},
  {Siltala}, {Silva}, {Smart}, {Smith}, {Solano}, {Solitro}, {Sordo}, {Soria
  Nieto}, {Souchay}, {Spagna}, {Spoto}, {Stampa}, {Steele},
  {Steidelm{\"u}ller}, {Stephenson}, {Stoev}, {Suess}, {Surdej}, {Szabados},
  {Szegedi-Elek}, {Tapiador}, {Taris}, {Tauran}, {Taylor}, {Teixeira},
  {Terrett}, {Teyssand ier}, {Thuillot}, {Titarenko}, {Torra Clotet}, {Turon},
  {Ulla}, {Utrilla}, {Uzzi}, {Vaillant}, {Valentini}, {Valette}, {van Elteren},
  {Van Hemelryck}, {van Leeuwen}, {Vaschetto}, {Vecchiato}, {Veljanoski},
  {Viala}, {Vicente}, {Vogt}, {von Essen}, {Voss}, {Votruba}, {Voutsinas},
  {Walmsley}, {Weiler}, {Wertz}, {Wevers}, {Wyrzykowski}, {Yoldas},
  {{\v{Z}}erjal}, {Ziaeepour}, {Zorec}, {Zschocke}, {Zucker}, {Zurbach}, \&
  {Zwitter}}]{gaia18}
{Gaia Collaboration}, {Brown}, A.~G.~A., {Vallenari}, A., {et~al.} 2018, \aap,
  616, A1

\bibitem[{{Grady} {et~al.}(2018){Grady}, {Brown}, {Welsh}, {Roberge}, {Kamp},
  \& {Rivi{\`e}re Marichalar}}]{grady18_hd172555}
{Grady}, C.~A., {Brown}, A., {Welsh}, B., {et~al.} 2018, \aj, 155, 242

\bibitem[{{Herter} {et~al.}(2012){Herter}, {Adams}, {De Buizer}, {Gull},
  {Schoenwald}, {Henderson}, {Keller}, {Nikola}, {Stacey}, \&
  {Vacca}}]{herter12}
{Herter}, T.~L., {Adams}, J.~D., {De Buizer}, J.~M., {et~al.} 2012, \apjl, 749,
  L18

\bibitem[{{Holden}(1976)}]{holden76}
{Holden}, F. 1976, \pasp, 88, 52

\bibitem[{{Hughes} {et~al.}(2018){Hughes}, {Duch{\^e}ne}, \&
  {Matthews}}]{hughes18}
{Hughes}, A.~M., {Duch{\^e}ne}, G., \& {Matthews}, B.~C. 2018, \araa, 56, 541

\bibitem[{{Jackson} \& {Wyatt}(2012)}]{jackson12}
{Jackson}, A.~P., \& {Wyatt}, M.~C. 2012, MNRAS, 425, 657

\bibitem[{{Jackson} {et~al.}(2014){Jackson}, {Wyatt}, {Bonsor}, \&
  {Veras}}]{jackson14}
{Jackson}, A.~P., {Wyatt}, M.~C., {Bonsor}, A., \& {Veras}, D. 2014, MNRAS,
  440, 3757

\bibitem[{{Johnson} \& {Melosh}(2012)}]{johnson12a}
{Johnson}, B.~C., \& {Melosh}, H.~J. 2012, Icarus, 217, 416

\bibitem[{{Johnson} {et~al.}(2012){Johnson}, {Lisse}, {Chen}, {Melosh},
  {Wyatt}, {Thebault}, {Henning}, {Gaidos}, {Elkins-Tanton}, {Bridges}, \&
  {Morlok}}]{johnson12b}
{Johnson}, B.~C., {Lisse}, C.~M., {Chen}, C.~H., {et~al.} 2012, ApJ, 761, 45

\bibitem[{{Kataza} {et~al.}(2000){Kataza}, {Okamoto}, {Takubo}, {Onaka},
  {Sako}, {Nakamura}, {Miyata}, \& {Yamashita}}]{kataza00_comics}
{Kataza}, H., {Okamoto}, Y., {Takubo}, S., {et~al.} 2000, Society of
  Photo-Optical Instrumentation Engineers (SPIE) Conference Series, Vol. 4008,
  {COMICS: the cooled mid-infrared camera and spectrometer for the Subaru
  telescope}, ed. M.~{Iye} \& A.~F. {Moorwood}, 1144--1152

\bibitem[{{Kemper} {et~al.}(2004){Kemper}, {Vriend}, \& {Tielens}}]{kemper04}
{Kemper}, F., {Vriend}, W.~J., \& {Tielens}, A.~G.~G.~M. 2004, \apj, 609, 826

\bibitem[{{Kennedy} \& {Wyatt}(2013)}]{kennedy_wyatt13}
{Kennedy}, G.~M., \& {Wyatt}, M.~C. 2013, MNRAS, 433, 2334

\bibitem[{{Kessler-Silacci} {et~al.}(2006){Kessler-Silacci}, {Augereau},
  {Dullemond}, {Geers}, {Lahuis}, {Evans}, {van Dishoeck}, {Blake}, {Boogert},
  {Brown}, {J{\o}rgensen}, {Knez}, \& {Pontoppidan}}]{kessler06_c2d}
{Kessler-Silacci}, J., {Augereau}, J.-C., {Dullemond}, C.~P., {et~al.} 2006,
  \apj, 639, 275

\bibitem[{{Kiefer} {et~al.}(2014){Kiefer}, {Lecavelier des Etangs}, {Augereau},
  {Vidal-Madjar}, {Lagrange}, \& {Beust}}]{kiefer14}
{Kiefer}, F., {Lecavelier des Etangs}, A., {Augereau}, J.~C., {et~al.} 2014,
  \aap, 561, L10

\bibitem[{{Kochanek} {et~al.}(2017){Kochanek}, {Shappee}, {Stanek}, {Holoien},
  {Thompson}, {Prieto}, {Dong}, {Shields}, {Will}, {Britt}, {Perzanowski}, \&
  {Pojma{\'n}ski}}]{kochanek17}
{Kochanek}, C.~S., {Shappee}, B.~J., {Stanek}, K.~Z., {et~al.} 2017, PASP, 129,
  104502

\bibitem[{{Koike} {et~al.}(2013){Koike}, {Noguchi}, {Chihara}, {Suto},
  {Ohtaka}, {Imai}, {Matsumoto}, \& {Tsuchiyama}}]{koike13}
{Koike}, C., {Noguchi}, R., {Chihara}, H., {et~al.} 2013, \apj, 778, 60

\bibitem[{{Kral} {et~al.}(2015){Kral}, {Th{\'e}bault}, {Augereau},
  {Boccaletti}, \& {Charnoz}}]{kral15}
{Kral}, Q., {Th{\'e}bault}, P., {Augereau}, J.-C., {Boccaletti}, A., \&
  {Charnoz}, S. 2015, A\&A, 573, A39

\bibitem[{{Krivov}(2010)}]{krivov10}
{Krivov}, A.~V. 2010, Research in Astronomy and Astrophysics, 10, 383

\bibitem[{{Lauretta} \& {McSween}(2006)}]{lauretta06}
{Lauretta}, D.~S., \& {McSween}, H.~Y. 2006, {Meteorites and the Early Solar
  System II}

\bibitem[{{Lawler} {et~al.}(2009){Lawler}, {Beichman}, {Bryden}, {Ciardi},
  {Tanner}, {Su}, {Stapelfeldt}, {Lisse}, \& {Harker}}]{lawler09}
{Lawler}, S.~M., {Beichman}, C.~A., {Bryden}, G., {et~al.} 2009, ApJ, 705, 89

\bibitem[{{Lebouteiller} {et~al.}(2011){Lebouteiller}, {Barry}, {Spoon},
  {Bernard-Salas}, {Sloan}, {Houck}, \& {Weedman}}]{cassis_ref}
{Lebouteiller}, V., {Barry}, D.~J., {Spoon}, H.~W.~W., {et~al.} 2011, \apjs,
  196, 8

\bibitem[{{Lieman-Sifry} {et~al.}(2016){Lieman-Sifry}, {Hughes}, {Carpenter},
  {Gorti}, {Hales}, \& {Flaherty}}]{lieman-sifry16}
{Lieman-Sifry}, J., {Hughes}, A.~M., {Carpenter}, J.~M., {et~al.} 2016, \apj,
  828, 25

\bibitem[{{Liou} \& {Malhotra}(1997)}]{liou_malhotra97}
{Liou}, J.~C., \& {Malhotra}, R. 1997, Science, 275, 375

\bibitem[{{Lisse} {et~al.}(2008){Lisse}, {Chen}, {Wyatt}, \&
  {Morlok}}]{lisse08}
{Lisse}, C.~M., {Chen}, C.~H., {Wyatt}, M.~C., \& {Morlok}, A. 2008, ApJ, 673,
  1106

\bibitem[{{Lisse} {et~al.}(2009){Lisse}, {Chen}, {Wyatt}, {Morlok}, {Song},
  {Bryden}, \& {Sheehan}}]{lisse09}
{Lisse}, C.~M., {Chen}, C.~H., {Wyatt}, M.~C., {et~al.} 2009, ApJ, 701, 2019

\bibitem[{{Lisse} {et~al.}(2020){Lisse}, {Meng}, {Sitko}, {Morlok}, {Johnson},
  {Jackson}, {Vervack}, {Chen}, {Wolk}, {Lucas}, {Marengo}, \&
  {Britt}}]{badlisse2020}
{Lisse}, C.~M., {Meng}, H.~Y.~A., {Sitko}, M.~L., {et~al.} 2020, arXiv
  e-prints, arXiv:2003.06870

\bibitem[{{MacGregor} {et~al.}(2020){MacGregor}, {Osten}, \&
  {Hughes}}]{macgregor20}
{MacGregor}, A.~M., {Osten}, R.~A., \& {Hughes}, A.~M. 2020, \apj, 891, 80

\bibitem[{{Mamajek} \& {Bell}(2014)}]{mamajek_bell14}
{Mamajek}, E.~E., \& {Bell}, C. P.~M. 2014, \mnras, 445, 2169

\bibitem[{{Marton} {et~al.}(2017){Marton}, {Calzoletti}, {Perez Garcia},
  {Kiss}, {Paladini}, {Altieri}, {Sanchez Portal}, {Kidger}, \& {the Herschel
  Point Source Catalogue Working Group}}]{marton17}
{Marton}, G., {Calzoletti}, L., {Perez Garcia}, A.~M., {et~al.} 2017, arXiv
  e-prints, arXiv:1705.05693

\bibitem[{{Matthews} {et~al.}(2014){Matthews}, {Krivov}, {Wyatt}, {Bryden}, \&
  {Eiroa}}]{matthews14}
{Matthews}, B.~C., {Krivov}, A.~V., {Wyatt}, M.~C., {Bryden}, G., \& {Eiroa},
  C. 2014, in Protostars and Planets VI, ed. H.~{Beuther}, R.~S. {Klessen},
  C.~P. {Dullemond}, \& T.~{Henning}, 521

\bibitem[{{Melis}(2016)}]{melis16}
{Melis}, C. 2016, in IAU Symposium, Vol. 314, Young Stars \&amp; Planets Near
  the Sun, ed. J.~H. {Kastner}, B.~{Stelzer}, \& S.~A. {Metchev}, 241--246

\bibitem[{{Melis} {et~al.}(2010){Melis}, {Zuckerman}, {Rhee}, \&
  {Song}}]{melis10}
{Melis}, C., {Zuckerman}, B., {Rhee}, J.~H., \& {Song}, I. 2010, ApJL, 717, L57

\bibitem[{{Melis} {et~al.}(2012){Melis}, {Zuckerman}, {Rhee}, {Song}, {Murphy},
  \& {Bessell}}]{melis12}
{Melis}, C., {Zuckerman}, B., {Rhee}, J.~H., {et~al.} 2012, Nature, 487, 74

\bibitem[{{Melis} {et~al.}(2013){Melis}, {Zuckerman}, {Rhee}, {Song}, {Murphy},
  \& {Bessell}}]{melis13}
---. 2013, ApJ, 778, 12

\bibitem[{{Meng} {et~al.}(2017){Meng}, {Rieke}, {Su}, \&
  {G{\'a}sp{\'a}r}}]{meng17}
{Meng}, H.~Y.~A., {Rieke}, G.~H., {Su}, K.~Y.~L., \& {G{\'a}sp{\'a}r}, A. 2017,
  ApJ, 836, 34

\bibitem[{{Meng} {et~al.}(2014){Meng}, {Su}, {Rieke}, {Stevenson}, {Plavchan},
  {Rujopakarn}, {Lisse}, {Poshyachinda}, \& {Reichart}}]{meng14}
{Meng}, H.~Y.~A., {Su}, K.~Y.~L., {Rieke}, G.~H., {et~al.} 2014, Science, 345,
  1032

\bibitem[{{Meng} {et~al.}(2015){Meng}, {Su}, {Rieke}, {Rujopakarn}, {Myers},
  {Cook}, {Erdelyi}, {Maloney}, {McMath}, {Persha}, {Poshyachinda}, \&
  {Reichart}}]{meng15}
---. 2015, ApJ, 805, 77

\bibitem[{{Meyer} {et~al.}(2001){Meyer}, {Backman}, {Mamajek}, {Herrera},
  {Hinz}, {Carpenter}, {Hoffman}, \& {Hora}}]{meyer01}
{Meyer}, M.~R., {Backman}, D., {Mamajek}, E.~E., {et~al.} 2001, in American
  Astronomical Society Meeting Abstracts, Vol. 199, American Astronomical
  Society Meeting Abstracts, 76.08

\bibitem[{{Mittal} {et~al.}(2015){Mittal}, {Chen}, {Jang-Condell}, {Manoj},
  {Sargent}, {Watson}, \& {Lisse}}]{mittal15}
{Mittal}, T., {Chen}, C.~H., {Jang-Condell}, H., {et~al.} 2015, \apj, 798, 87

\bibitem[{{Morlok} {et~al.}(2014){Morlok}, {Mason}, {Anand}, {Lisse},
  {Bullock}, \& {Grady}}]{morlok14}
{Morlok}, A., {Mason}, A.~B., {Anand}, M., {et~al.} 2014, Icarus, 239, 1

\bibitem[{{Okamoto} {et~al.}(2003){Okamoto}, {Kataza}, {Yamashita}, {Miyata},
  {Sako}, {Takubo}, {Honda}, \& {Onaka}}]{okamoto03_comics}
{Okamoto}, Y.~K., {Kataza}, H., {Yamashita}, T., {et~al.} 2003, Society of
  Photo-Optical Instrumentation Engineers (SPIE) Conference Series, Vol. 4841,
  {Improved performances and capabilities of the Cooled Mid-Infrared Camera and
  Spectrometer (COMICS) for the Subaru Telesc ope}, ed. M.~{Iye} \& A.~F.~M.
  {Moorwood}, 169--180

\bibitem[{{Olofsson} {et~al.}(2013){Olofsson}, {Henning}, {Nielbock},
  {Augereau}, {Juh{\`a}sz}, {Oliveira}, {Absil}, \& {Tamanai}}]{olofsson13}
{Olofsson}, J., {Henning}, T., {Nielbock}, M., {et~al.} 2013, A\&A, 551, A134

\bibitem[{{Olofsson} {et~al.}(2012){Olofsson}, {Juh{\'a}sz}, {Henning},
  {Mutschke}, {Tamanai}, {Mo{\'o}r}, \& {{\'A}brah{\'a}m}}]{olofsson12}
{Olofsson}, J., {Juh{\'a}sz}, A., {Henning}, T., {et~al.} 2012, A\&A, 542, A90

\bibitem[{{Osten} {et~al.}(2013){Osten}, {Livio}, {Lubow}, {Pringle},
  {Soderblom}, \& {Valenti}}]{osten13}
{Osten}, R., {Livio}, M., {Lubow}, S., {et~al.} 2013, ApJL, 765, L44

\bibitem[{{Pecaut} {et~al.}(2012){Pecaut}, {Mamajek}, \& {Bubar}}]{pecaut12}
{Pecaut}, M.~J., {Mamajek}, E.~E., \& {Bubar}, E.~J. 2012, \apj, 746, 154

\bibitem[{{Rebull} {et~al.}(2008){Rebull}, {Stapelfeldt}, {Werner}, {Mannings},
  {Chen}, {Stauffer}, {Smith}, {Song}, {Hines}, \& {Low}}]{rebull08}
{Rebull}, L.~M., {Stapelfeldt}, K.~R., {Werner}, M.~W., {et~al.} 2008, ApJ,
  681, 1484

\bibitem[{{Rhee} {et~al.}(2008){Rhee}, {Song}, \& {Zuckerman}}]{rhee08}
{Rhee}, J.~H., {Song}, I., \& {Zuckerman}, B. 2008, ApJ, 675, 777

\bibitem[{{Rieke} {et~al.}(2008){Rieke}, {Blaylock}, {Decin}, {Engelbracht},
  {Ogle}, {Avrett}, {Carpenter}, {Cutri}, {Armus}, {Gordon}, {Gray}, {Hinz},
  {Su}, \& {Willmer}}]{rieke08}
{Rieke}, G.~H., {Blaylock}, M., {Decin}, L., {et~al.} 2008, AJ, 135, 2245

\bibitem[{{Riviere-Marichalar} {et~al.}(2014){Riviere-Marichalar}, {Barrado},
  {Montesinos}, {Duch{\^e}ne}, {Bouy}, {Pinte}, {Menard}, {Donaldson}, {Eiroa},
  {Krivov}, {Kamp}, {Mendigut{\'\i}a}, {Dent}, \&
  {Lillo-Box}}]{riviere-marichalar14}
{Riviere-Marichalar}, P., {Barrado}, D., {Montesinos}, B., {et~al.} 2014, \aap,
  565, A68

\bibitem[{{Sargent} {et~al.}(2009){Sargent}, {Forrest}, {Tayrien}, {McClure},
  {Watson}, {Sloan}, {Li}, {Manoj}, {Bohac}, {Furlan}, {Kim}, \&
  {Green}}]{sargent09}
{Sargent}, B.~A., {Forrest}, W.~J., {Tayrien}, C., {et~al.} 2009, \apjs, 182,
  477

\bibitem[{{Sch{\"u}tz} {et~al.}(2005){Sch{\"u}tz}, {Meeus}, \&
  {Sterzik}}]{schutz05}
{Sch{\"u}tz}, O., {Meeus}, G., \& {Sterzik}, M.~F. 2005, \aap, 431, 175

\bibitem[{{Sierchio} {et~al.}(2014){Sierchio}, {Rieke}, {Su}, \&
  {G{\'a}sp{\'a}r}}]{sierchio14}
{Sierchio}, J.~M., {Rieke}, G.~H., {Su}, K.~Y.~L., \& {G{\'a}sp{\'a}r}, A.
  2014, ApJ, 785, 33

\bibitem[{{Smith} {et~al.}(2012){Smith}, {Wyatt}, \& {Haniff}}]{smith12}
{Smith}, R., {Wyatt}, M.~C., \& {Haniff}, C.~A. 2012, \mnras, 422, 2560

\bibitem[{{Song} {et~al.}(2005){Song}, {Zuckerman}, {Weinberger}, \&
  {Becklin}}]{song05}
{Song}, I., {Zuckerman}, B., {Weinberger}, A.~J., \& {Becklin}, E.~E. 2005,
  Nature, 436, 363

\bibitem[{{Su} {et~al.}(2015){Su}, {Morrison}, {Malhotra}, {Smith}, {Balog}, \&
  {Rieke}}]{su15}
{Su}, K. Y.~L., {Morrison}, S., {Malhotra}, R., {et~al.} 2015, \apj, 799, 146

\bibitem[{{Su} \& {Rieke}(2014)}]{su14}
{Su}, K.~Y.~L., \& {Rieke}, G.~H. 2014, in IAU Symposium, Vol. 299, IAU
  Symposium, ed. M.~{Booth}, B.~C. {Matthews}, \& J.~R. {Graham}, 318--321

\bibitem[{{Su} {et~al.}(2019){Su}, {Jackson}, {G{\'a}sp{\'a}r}, {Rieke},
  {Dong}, {Olofsson}, {Kennedy}, {Leinhardt}, {Malhotra}, {Hammer}, {Meng},
  {Rujopakarn}, {Rodriguez}, {Pepper}, {Reichart}, {James}, \&
  {Stassun}}]{su19}
{Su}, K.~Y.~L., {Jackson}, A.~P., {G{\'a}sp{\'a}r}, A., {et~al.} 2019, AJ, 157,
  202

\bibitem[{{Thebault} \& {Kral}(2019)}]{thebault19}
{Thebault}, P., \& {Kral}, Q. 2019, \aap, 626, A24

\bibitem[{{Thompson} {et~al.}(2019){Thompson}, {Weinberger}, {Keller},
  {Arnold}, \& {Stark}}]{thompson19}
{Thompson}, M.~A., {Weinberger}, A.~J., {Keller}, L.~D., {Arnold}, J.~A., \&
  {Stark}, C.~C. 2019, \apj, 875, 45

\bibitem[{{Torres} {et~al.}(2006){Torres}, {Quast}, {da Silva}, {de La Reza},
  {Melo}, \& {Sterzik}}]{torres06}
{Torres}, C.~A.~O., {Quast}, G.~R., {da Silva}, L., {et~al.} 2006, \aap, 460,
  695

\bibitem[{{van Dishoeck}(2004)}]{vandishoeck04}
{van Dishoeck}, E.~F. 2004, \araa, 42, 119

\bibitem[{{Weinberger} {et~al.}(2011){Weinberger}, {Becklin}, {Song}, \&
  {Zuckerman}}]{weinberger11}
{Weinberger}, A.~J., {Becklin}, E.~E., {Song}, I., \& {Zuckerman}, B. 2011,
  \apj, 726, 72

\bibitem[{{Wyatt}(2008)}]{wyatt08}
{Wyatt}, M.~C. 2008, Astron. Astrophys. Rev., 46, 339

\bibitem[{{Wyatt} {et~al.}(2007){Wyatt}, {Smith}, {Greaves}, {Beichman},
  {Bryden}, \& {Lisse}}]{wyatt07a}
{Wyatt}, M.~C., {Smith}, R., {Greaves}, J.~S., {et~al.} 2007, ApJ, 658, 569

\bibitem[{{Zellner}(2019)}]{zellner19}
{Zellner}, N.~E.~B. 2019, Journal of Geophysical Research (Planets), 124, 2686

\end{thebibliography}
\end{document}